\pdfoutput=1
\documentclass[a4paper,11pt]{article}
\usepackage{jheppub} 
\usepackage{amssymb}
\usepackage{amsmath}
\usepackage{amsthm}
\usepackage{dsfont}
\usepackage{hyperref}
\usepackage{color}
\usepackage{graphicx}
\usepackage{multirow}
\usepackage[utf8]{inputenc}

\newcommand{\beq}{\begin{equation}}
\newcommand{\eeq}{\end{equation}}

\def\[{\left[}
\def\]{\right]}
\def\({\left(}
\def\){\right)}
\def\[{\left[}

\def\f{f}
\def\Qy{P}
\def\b{\beta}

\def\z{s}
\def\d{b}
\def\T{\mathcal{T}}
\def\Q{\mathcal{Q}}

\def\M{\mathcal{M}}
\def\F{\mathcal{F}}
\def\A{\mathcal{A}}
\def\B{\mathcal{B}}
\def\M{\mathcal{M}}
\def\N{\mathcal{N}}
\def\W{\mathcal{W}}
\def\Y{\mathcal{Y}}
\def\D{\mathcal{D}}
\def\Z{\mathcal{Z}}

\def\U{C^{+}}
\def\E{\tilde{E}}
\def\su{\mathfrak{su}}
\def\sl{\mathfrak{sl}}
\def\psu{\mathfrak{su}}

\def\p{\tilde{p}}

\def\am{a^{\textrm{max}}}
\def\rm{r_{\textrm{min}}}

\title{\Large \bf Three-point functions at strong coupling in the BMN limit}

\author{Benjamin Basso}
\author{and De-liang Zhong}
\affiliation{Laboratoire de physique de l'Ecole normale sup\'erieure,
ENS, Universit\'e PSL, CNRS, Sorbonne Universit\'e, Universit\'e Paris-Diderot, Sorbonne Paris Cit\'e,
24 rue Lhomond, 75005 Paris, France}

\abstract{We consider structure constants of single-trace operators at strong coupling in planar $\mathcal{N}=4$ SYM theory using the hexagon formalism. We concentrate on heavy-heavy-light correlators where the heavy operators are BMN operators, with large R-charges and finite anomalous dimensions, and the light one is a finite-charge chiral primary operator. They describe the couplings between two highly boosted strings and a supergravity mode in the bulk dual. In the hexagon framework, two sums over virtual magnons are needed to bind the hexagons together around the light operator. We evaluate these sums explicitly at strong coupling, for a certain choice of BMN operators, and show that they factorise into a ratio of Gamma functions and a simple stringy prefactor. The former originates from giant mirror magnons scanning the AdS geometry while the latter stems from small fluctuations around the BMN vacuum. The resulting structure constants have poles at positions where an enhanced mixing with double-trace operators is expected and zeros whenever the process is forbidden by supersymmetry. We also discuss the transition to the classical regime, when the length of the light operator scales like the string tension, where we observe similitudes with the Neumann coefficients of the pp-wave String Field Theory vertex.}

\begin{document}
\maketitle

\renewcommand{\thefootnote}{\fnsymbol{footnote}}

\renewcommand{\thefootnote}{\arabic{footnote}}

\section{Introduction}\label{Intro}

There has been a great deal of activity recently regarding correlation functions of local operators in planar $\mathcal{N} = 4$ SYM and in its holographic dual, IIB superstring theory in $AdS_5\times S^5$. On the one hand, building on Mellin space techniques \cite{Penedones:2010ue,Mack:2009mi} and bootstrap ideas, new approaches have been developed \cite{Rastelli:2016nze,Alday:2017xua,Aprile:2017bgs,Alday:2017vkk,Aprile:2018efk,Caron-Huot:2018kta,Binder:2019jwn,Goncalves:2019znr} to deal more efficiently with the supergravity regime, corresponding to the strong coupling limit, $g^2 = \lambda/(4\pi)^2 \gg 1$, of the large-$N_{c}$ gauge theory. They led to spectacular results, starting with a conjecture \cite{Rastelli:2016nze} for the $1/N_{c}^2$ correction to the 4pt functions of single-trace chiral primary operators of arbitrary dimensions $\sim g^0$, which generalises earlier results and proposals, see \cite{Dolan:2006ec} and references therein. Further considerations unveiled hidden symmetries of the supergravity regime \cite{Caron-Huot:2018kta,Aprile:2018efk} and yielded lots of new OPE data for double-trace operators at strong coupling \cite{Alday:2017xua,Aprile:2017bgs,Aprile:2017xsp,Alday:2017vkk,Aprile:2018efk}. They suggest the exciting possibility that more general correlators can be found in the supergravity regime without ever using a single Witten diagram.

On the other hand, in a different vein, the integrability technology, see \cite{Beisert:2010jr} for a review, fostered the development of form-factor methods aiming at solving correlation functions, or scattering amplitudes, for any $g$ in the large $N_{c}$ limit \cite{Escobedo:2010xs,Basso:2013vsa,Bajnok:2015hla,Basso:2015zoa,Fleury:2016ykk,Eden:2016xvg,Bargheer:2017nne,Eden:2017ozn,Ben-Israel:2018ckc}. Among these techniques, the hexagon method appears as the most versatile. Developed initially for the 3pt functions \cite{Basso:2015zoa}, it has been extended such as to cover higher-point functions \cite{Fleury:2016ykk,Eden:2016xvg} and non-planar corrections \cite{Bargheer:2017nne,Eden:2017ozn}. (See also~\cite{Jiang:2019xdz,Kim:2019gcq} for recent applications to integrable defects.) The method passed all the tests at weak coupling, see~\cite{Fleury:2017eph,Chicherin:2018avq,Bargheer:2018jvq,Coronado:2018ypq,Coronado:2018cxj,Kostov:2019stn,Kostov:2019auq} for recent examples, and has been checked at strong coupling as well, although to a lesser extent, in the semiclassical regime \cite{Jiang:2016ulr} corresponding to minimal surfaces in $AdS_{5} \times S^{5}$~\cite{Kazama:2016cfl,Kazama:2013qsa,Kazama:2011cp,Janik:2011bd}. However, to date, the striking simplicity of the supergravity limit is still evading it.

In this paper we take a step towards the low-energy regime and apply the hexagon method at strong coupling to 3pt functions of single-trace operators involving one \textit{light} chiral primary operator, dual to a supergravity mode, and two \textit{heavy} operators dual to highly boosted strings. The latter are the standard BMN operators, carrying a large R charge and a finite anomalous dimension $\gamma$ and mapping to states with finitely many magnons moving on a very large spin chain. For simplicity, we will take one of the two states to be BPS, corresponding to the spin-chain supersymmetric vacuum. The 3pt functions of interest are thus the familiar ones, with two BPS and one non-BPS single-trace operators, $\mathcal{O}_{1,2}$ and $\mathcal{O}_{\gamma}$, of lengths $L_{1,2}$ and $L$, respectively,
\beq\label{C123}
C^{\circ \circ \bullet} = \left<\mathcal{O}_{1}(\infty)\mathcal{O}_{2}(1)\mathcal{O}_{\gamma}(0)\right>\, ,
\eeq
and with $L\sim L_{1} \sim g \gg 1$ and $L_{2}, \gamma \sim g^0$ in the heavy-heavy-light (HHL) kinematics. This set-up is interesting in that it enables to probe correlators at low energy and still avoids bottlenecks of the hexagon approach.

\begin{figure}
\begin{center}
\includegraphics[scale=0.35]{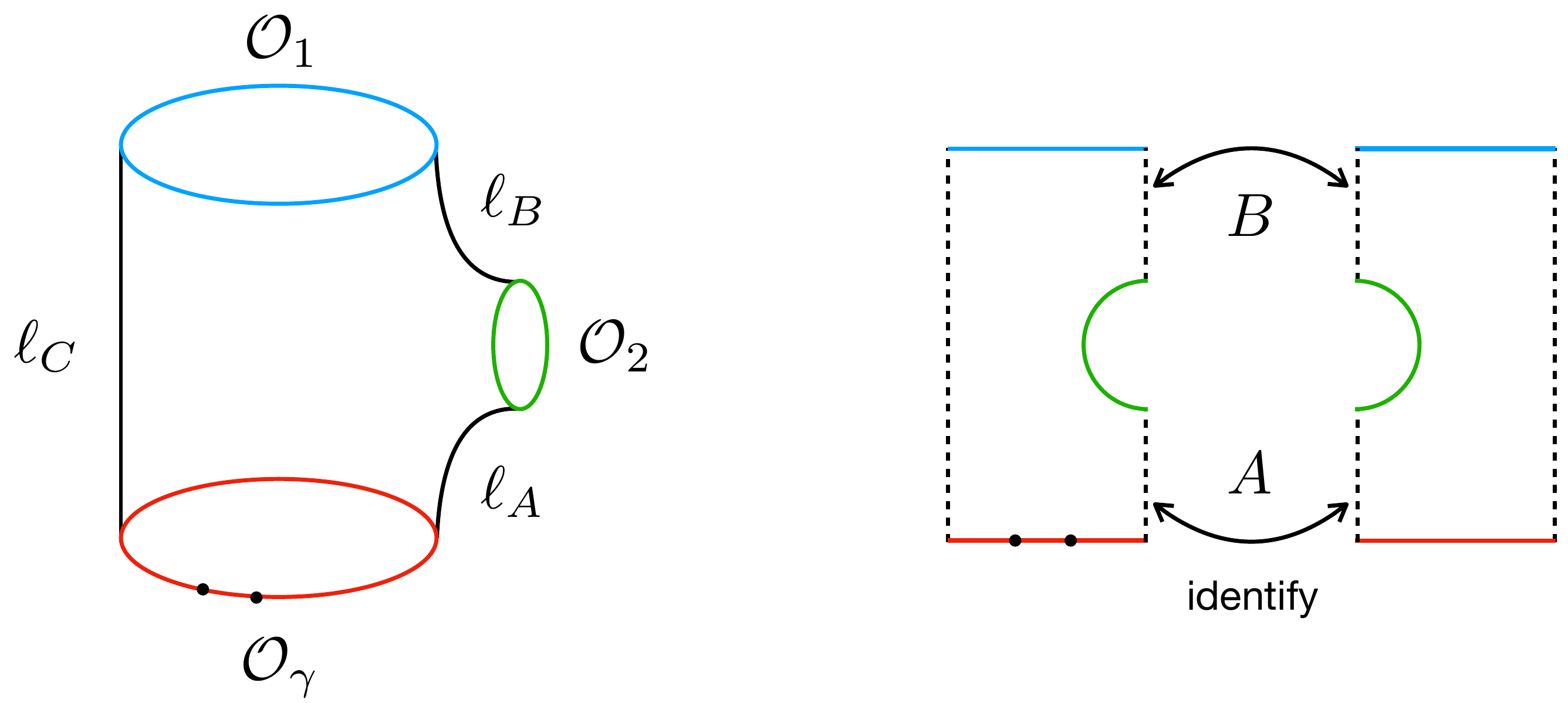}
\end{center}
\caption{Left panel: The pair-of-pant diagram representing the structure constant for single-trace operators of lengths $L_{1}, L_{2}$ and $L$. The geometry is characterised by three bridges, of thickness $\ell_{A, B, C}$, representing, and counting, the tree-level contractions among the operators. Cut opening along the bridges separates the hexagons. The dots in the picture are the excitations -magnons- of the bottom operator, with non-zero anomalous dimension $\gamma$; the other states are BPS. Right panel: The HHL limit $\ell_C \rightarrow \infty$ corresponds to decompactifying the cylinder along edge $C$ while keeping fixed the middle -light- operator with length $L_{2}$. The vertex can be obtained by gluing the hexagons along edges $A$ and $B$, using a complete sum of mirror magnons on each edge.}\label{vertexi} 
\end{figure}

To understand this point, recall that the idea is to build the string vertex by attaching two hexagons together along the seams of the pair-of-pants diagram, as shown in figure~\ref{vertexi}. The picture gets more quantitative at weak coupling where the spin-chain description takes over \cite{Escobedo:2010xs}. Each seam is then identified with a bridge of planar contractions among the spin-chain sites and acquires a thickness or length. The hexagons fully decouple when the three bridge lengths in the problem ($\ell_{A, B, C}$) are asymptotically large, which means much larger than $g$ at strong coupling. This requires in particular that all three operators carry extremely large charges and dimensions.

For smaller $\ell$'s a sum over a complete basis of virtual excitations, which move across the seams, must be included. These excitations - dubbed mirror magnons - encode the finite-size effects of the 3pt function geometry and computing their sum is a difficult task in general. It becomes unwieldy in the finite-length regime, which maps to a short-distance limit for the hexagon form factor series, and it looks almost hopeless when the mirror magnons are given the freedom to move across many bridges.

The HHL regime corresponds to $\ell_{A, B} \ll \ell_{C}$ and it minimises the problem by confining the mirror magnons to the neighbourhood of the light operator, $\ell_{A}+\ell_{B} = L_{2}$. Importantly, it prevents the mirror magnons from winding around the unprotected operator.%
\footnote{More precisely, we need $\ell_{C} \gg g$ to kill the finite-size effects along the bridge $C$, implying that the heavy operators have lengths $\gg g$. This approximation is also needed to keep the anomalous dimension of order $\sim g^0$ and to avoid the extremal points, see Section \ref{Sect5}. The 3pt functions are singular at these points, see~\cite{Freedman:1998tz,DHoker:1999jke,Alday:2013cwa,Minahan:2014usa,Korchemsky:2015cyx,Alday:2016mxe} for examples and discussions, and so are the mirror sums, which must be analytically continued.}
The latter are source of spurious divergences and require a dedicated treatment, which has not been fully worked out, see \cite{Basso:2017muf,Basso:2018cvy,Bajnok:2017mdf,Bajnok:2018fzx,Bajnok:2019cdf} for attempts and related discussions.

 The HHL kinematics was explored using worldsheet techniques in~\cite{Zarembo:2010rr,Costa:2010rz,Bajnok:2014sza,Bajnok:2016xxu,Bajnok:2015ftj,Bajnok:2015hla,Bajnok:2017mdf}. In particular, Ref.~\cite{Bajnok:2017mdf} studied a similar open-open-closed-string vertex, although in a slightly different regime, by wrapping an octagon around a closed-string operator. The hexagon picture is obtained by cutting smaller and thinking of the octagon as resulting from the gluing of two hexagons, as shown in the right panel of figure \ref{vertexi}. Two mirror sums are needed here, for the two bridges that stay finite,
\beq\label{bridges}
\ell_{A,B} = \frac{L_{2} \pm L \mp L_{1}}{2} = O(1)\, ,
\eeq
when $\ell_{C} = \frac{1}{2}(L_{1}+ L - L_{2}) \rightarrow \infty$. In this paper, we shall calculate them exactly when $\gamma\sim g^0$, using among other things the Pfaffian formula \cite{Basso:2017khq} and the associated summation techniques applied recently in \cite{Kostov:2019stn,Kostov:2019auq} for the 4pt functions. It will allow us to show that the 3pt functions factorise into ratios of Gamma functions and simple stringy prefactors, in line with Witten diagrams \cite{Freedman:1998tz,Aharony:1999ti} and with the pp-wave holographic dictionary \cite{Dobashi:2004nm,Dobashi:2004ka,Lee:2004cq}.

For illustration, when all the magnons on the unprotected operator spin in AdS, that is for~$\mathcal{O}_{L, \gamma} \sim \textrm{tr}\, \D^{M}\Z^{L}$ with $\Z$ a complex scalar field and $\D$ a lightcone derivative, we will get
\beq\label{intro}
C^{\circ\circ\bullet}/C^{\circ\circ\circ} = \M\times \frac{\Gamma(\ell_{B}-\tfrac{1}{2}\gamma)\Gamma(\ell_{A}+M+\tfrac{1}{2}\gamma)}{\Gamma(\ell_{B})\Gamma(\ell_{A})}\, ,
\eeq
with $C^{\circ\circ\circ}$ the structure constant for the chiral primary operator ($M, \gamma = 0$).%
\footnote{$C^{\circ\circ\circ}= \sqrt{L_{1}L_{2}L}/N_{c}$ at large $N_{c}$ \cite{Lee:1998bxa}.} A similar expression will be found for a larger family of operators carrying an additional spin along the sphere. The prefactor $\M$ will be common to all of them and expressed in terms of the BMN energies of the magnons, $\M = \prod_{i=1}^{M}(LE_{i})^{-1/2}$ with $\sum_{i}E_{i} = M+\gamma$. Combining insights from the semiclassical string results \cite{Kazama:2016cfl} and the hexagon representation, we will argue that these formulae are free from wrapping corrections and stay valid when $\ell_{C}\sim g$ as long as $L_{2} \ll \ell_{C}$ and $\gamma = O(1)$.

Finally, we will explore the transition to the classical regime $L_{2}\sim g$ where the Gamma functions give way to factorised dressing factors. In this limit we will be able to carry out a comparison with the Neumann coefficients of the pp-wave String Field Theory (SFT) vertex \cite{Bajnok:2017mdf,Bajnok:2015hla,Lucietti:2004wy,He:2002zu,Dobashi:2004nm,Dobashi:2004ka,Lee:2004cq}.%
\footnote{This vertex describes the near-collinear splitting of a string and is associated to geometries in which a bridge length is much smaller than the others, e.g., $\ell_{A} \ll \ell_{B, C} \sim g$.}

The plan of the paper is the following. In Section \ref{Sect2} we set up our notations, introduce the spin-chain states of interest and recall the main hexagon formulae for their structure constants. In Section \ref{Sect3} we analyse the one-mirror-magnon integral at strong coupling for both classical and quantum bridges, corresponding to $\ell \sim g$ and $\ell\sim 1$, respectively. We argue that the tower of mirror bound states can be replaced by a continuum of states in the latter case and proceed with their integration. We generalise the analysis to any number of mirror particles in Section \ref{Sect4} using the Pfaffian representation for the hexagon form factors and compute the mirror series. In Section \ref{Sect5} we put all the ingredients together for the structure constants, discuss their main properties and argue for the absence of wrapping corrections in the HHL regime. We conclude in Section \ref{Sect6}. The appendices contain additional material for the bravest readers.

\section{Generalities}\label{Sect2}

In this section we recall the general hexagon formulae for structure constants in the HHL regime. To begin with, we make more precise the operators that we shall be considering.

\subsection{Operators}

In this paper we consider planar 3pt functions between two BPS operators and one non-BPS operator using the hexagon formalism. The latter are single-trace chiral primary operators built out of complex scalar fields. Without loss of generality, we take them as
\beq
\mathcal{O}_{1} \sim \textrm{tr}\, \bar{\Z}^{L_{1}}\, , \qquad \mathcal{O}_{2} \sim \textrm{tr}\, (\Z+\bar{\Z}+\Y-\bar{\Y})^{L_{2}}\, , 
\eeq
where $\Z, \bar{\Z} = \phi_{1}\pm i\phi_{2}$ and $\Y, \bar{\Y} = \phi_{3}\pm i\phi_{4}$. The fused operator is a single-trace chain of length $L$, with a non-zero anomalous dimension $\gamma$. It reads as a spin-chain state on the vacuum $\textrm{tr}\, \Z^{L}$,
\beq\label{spins}
\mathcal{O}_{\gamma} \sim \textrm{tr}\, \Z\ldots \Z \chi_{1}\Z \ldots \Z\chi_{M} \Z \ldots \Z +\ldots\, ,
\eeq
with the extra fields $\chi_{i}$'s mapping to magnons. The dots indicate the need to smear the magnons, such as to obtain a conformal and R-symmetry primary.

Each magnon moves along the chain with a momentum $p$ and a corresponding energy
\beq\label{Ep}
E = \sqrt{1+16g^2 \sin^2{\frac{p}{2}}}\, .
\eeq
The total energy yields the anomalous dimension $\gamma$ of the operator, $\sum_{i=1}^{M}E_i = M +\gamma$, up to exponentially small corrections at $L\rightarrow \infty$. As usual with integrable models, the most useful variables are not the momenta but the rapidities that uniformize the interactions. In the case at hand, we get two rapidities $x^{\pm}$, related to each other $x^{\pm} = x(u\pm i/2)$ and to the more common Bethe rapidity $u$ through the Zhukowski map
\beq\label{xu}
x(u) = \frac{1}{2 g}(u+\sqrt{u^2-4g^2})\, .
\eeq
The dispersion relation can be written in these terms using
\beq\label{Egen}
E = 1 + 2i g (1/x^{+}-1/x^{-})\, , \qquad p = -i\log{(x^{+}/x^{-})}\, .
\eeq
At strong coupling the magnons can cover a wide range of energies, from $E\sim 1$ to $E\sim g$. The BMN operators correspond to the low-lying states in this spectrum; they carry $\gamma \sim 1$ and are composed of low-momentum modes $p_i\sim 1/g$.  (This domain is also known as the plane-wave region, as the magnon S matrix goes to 1 at strong coupling for such momenta.) As well known, and as one can see from (\ref{Ep}), this kinematics is relativistic. Relatedly, we can drop the $\pm$ superscripts in (\ref{Egen}) and proceed with a single Zhukowski variable, for each magnon,
\beq \label{eqn-xpmLargeG}
x_i^{\pm} = x_i \pm \frac{i x_i^2}{2g(x_i^2-1)} + O(1/g^2)\, ,
\eeq 
with $|x_i| > 1$, giving the relativistic spectrum in the form
\beq\label{Eipi}
\gamma = \sum_{i=1}^{M}\frac{2}{x_{i}^2-1}+O(1/g)\, , \qquad p_{i} = \frac{x_{i}}{g(x^2_{i}-1)} + O(1/g^2)\, .
\eeq
The variable $x$ relates then to the more common hyperbolic rapidity via $x = \textrm{coth}{(\tfrac{1}{2}\theta)}$.

In the large volume limit, $L\gg g$, there is no need for the quantization of the momenta. Hence, until we relax the latter assumption, in Section \ref{Sect5}, the rapidities $\{x_i\}$ will be treated as free parameters, as much as the energy $\gamma$. We shall find convenient however to impose the zero-momentum condition,
\beq\label{ZM}
0 = \sum_{i=1}^{M}p_{i} = \sum_{i=1}^{M}\frac{x_{i}}{g(x_{i}^2-1)}\, ,
\eeq
implementing the cyclicity of the state, as it will lead to further simplifications.

Besides the rapidity, each magnon carries a bi-fundamental index for its transformation property under the centrally extended $PSU(2|2)^2$ symmetry of the spin-chain vacuum \cite{Beisert:2006qh}. In this paper we shall restrict this index to the graded subspace generated by
\beq
\chi \in \left(\begin{array}{c} \phi_{1}\\ \psi_{1} \end{array}\right) \otimes \left(\begin{array}{c} \dot{\phi}_{1}\\ \dot{\psi}_{1} \end{array}\right) = \left(\begin{array}{cc} \mathcal{Y} & \bar{\Psi} \\ \Psi & \mathcal{D}\mathcal{Z}\end{array}\right)\, ,
\eeq
where $\Y$ and $\D$ are the scalar field and lightcone derivative introduced earlier, and $\Psi$ and $\bar{\Psi}$ are the gauginos they can mix with, $\mathcal{Y}\mathcal{D}\mathcal{Z} \sim \Psi \bar{\Psi}$.
The corresponding linear space of local operators is known as the $\psu(1,1|2)$ sector, see \cite{Beisert:2005fw}, and it is closed under renormalisation at any coupling.

We get rid of the flavour indices by building scattering eigenstates.%
\footnote{One could also work with indices in the decompactification limit. However, in the hexagon framework, the interactions on the pair-of-pants involve the magnon scattering amplitudes and their diagonalisation simplifies the algebra.}
The procedure is performed in the usual manner using the nested Bethe ansatz \cite{Beisert:2005tm}. We shall make use of the compact grading, treating the scalars as the main excitations and the rest as defining the nested layers, see \cite{Beisert:2005fw}. Later on, we will convert the expressions so-obtained to the alternative, non-compact grading, where the derivatives play the leading role.

Two nested levels are needed here, for the left- and the right-handed fermions, and two sets of auxiliary rapidities $\{y_i, i=1, \ldots , N\},\{\dot{y}_i, i = 1, \ldots \dot{N}\}$ are introduced.  At strong coupling, for scattering eigenstates, they obey the linearised Bethe ansatz equations  \cite{Beisert:2005fw}
\beq\label{BAE}
\sum_{j=1}^{M} \frac{x_{j}^2}{(x^2_{j}-1)(y- x_{j})} = 0\, , \qquad \forall y \in \{y_{i}\} \cup \{\dot{y}_{i}\}\, , 
\eeq
modulo terms that vanish for cyclic states (\ref{ZM}). We will not need to know much about the solutions to these equations; enough to note, numerically, that their modules are greater than $1$, for all $|x_{j}|>1$. The exception to the rule is the root $y=0$, which always exists for cyclic states.

For application to structure constants with two half-BPS operators we can fold the wave functions as only the diagonal states with $\{y_{i}\} = \{\dot{y}_i\}$ return a non-zero answer. This superselection rule was derived in \cite{Basso:2017khq} from the diagonal $PSU(2|2)$ Yangian symmetry of the hexagon form factors. It extends the global selection rule $N = \dot{N}$ which states that only left-right symmetric representations show up in the OPE of two chiral primary operators. Hence, summarising, the diagonal operators to be studied look like
\beq\label{fields}
\mathcal{O}_{\gamma} \sim \textrm{tr}\, \D^{S}\Z^{L-Y}\Y^{Y} \, ,
\eeq
up to mixing. They carry Lorentz spin $S = N$ and scalar charge $Y = M-N$, with $M,N$ the excitation numbers in the compact grading. 

\subsection{Hexagon sums} \label{sec:hexagonSum}

\begin{figure}
\begin{center}
\includegraphics[scale=0.35]{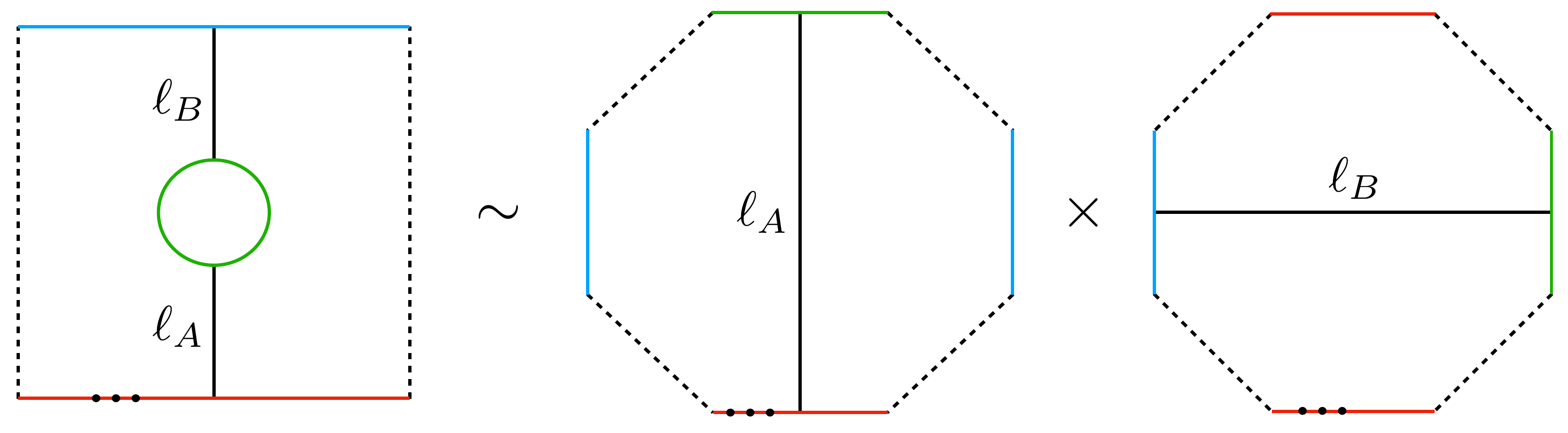}
\end{center}
\caption{The vertex is covered using two hexagons, which are stitched together along two edges of thickness $\ell_{A}$ and $\ell_{B}$. According to the hexagon formula, for a half-BPS insertion the structure constant factorises into two octagons, shown in the right panel. The dots stand for real magnons which are spectators at the boundary.}\label{vertex} 
\end{figure}

The hexagon construction allows us to obtain a representation for the HHL structure constants by attaching two hexagons around the light operator, as in figure \ref{vertexi}. The gluing is performed by inserting sums of mirror states on the identified edges. As said earlier, the hexagons generically develop divergences when they are wrapped around a local operator. Importantly, the operator surrounded here is half BPS and as such is protected from any such divergences. More than that, according to the hexagon formula worked out in \cite{Basso:2015eqa}, the mirror magnons on the two sides of the half-BPS operator do not talk to each other.%
\footnote{The analysis in \cite{Basso:2015eqa} was performed for Bethe states in rank-one sectors. Its higher-rank generalisation is straightforward following considerations in \cite{Basso:2017khq}.}
In other words, the structure constants factorise, see figure \ref{vertex},
\beq\label{factorise}
C^{\circ\circ\bullet}/C^{\circ\circ\circ} = \N_{L}\A(\ell_{A}) \B (\ell_{B})\, ,
\eeq
with $\A(\ell_{A})$ the result of gluing uniquely along the bridge $A$ and similarly for $\B(\ell_{B})$, and the problem boils down to studying each mirror sum separately. We should stress that although the mirror magnons on the two bridges do not see each other, they feel separately the presence of the real magnons at the boundary. In fact, it is these who make the difference between $\A$ and $\B$.

Let us now look closer at the various factors in (\ref{factorise}), following \cite{Basso:2015eqa} up to changes in the notations.

First, $\N_{L}$ is a simple factor, denoted as $\texttt{Gaudin}$ in \cite{Basso:2015eqa}, which accounts for the overall normalisation.%
\footnote{In particular, it takes into account the fact that the structure constants computed here are for operators that are canonically normalised.} It depends on the magnons' rapidities, $u_{i} = g(x_{i}+1/x_{i})$, and reads, in absolute value,
\beq\label{pre-factor}
|\N_{L}|^2 = \frac{1}{\mathcal{G}_{L}}\prod_{i=1}^{M}|\frac{\mu(u_{i})}{dp(u_{i})/du}| \times \prod_{i< j}^{M}\Delta(u_{i}, u_{j})\, ,
\eeq
where $\Delta(u_{i}, u_{j}) = h(u_{i}, u_{j})h(u_{j}, u_{i})$ is the squared module of the two-scalar hexagon form factor and $\mu$ is the hexagon measure. The former trivialises at strong coupling for (distinct) low-momentum magnons,
\beq\label{hD}
h, \Delta \rightarrow 1\, .
\eeq
The momentum-space measure is a simple factor, which takes the same form at any coupling,
\beq
|\mu(u) \frac{du}{dp}| = \frac{4g^2}{E(E^2-1)}\, ,
\eeq
when expressed in terms of the magnon energy, see Eq.~(\ref{Ep}). Lastly, $\mathcal{G}_{L}$ is the Gaudin determinant normalising the asymptotic wave function at large $L$.%
\footnote{Strictly speaking, when the nested levels are excited, $\mathcal{G}_{L}$ is the square of the Gaudin norm computed at fixed mode numbers for the auxiliary roots, see \cite{Basso:2017khq}. This distinction is irrelevant here since there is no interaction.} In the plane-wave regime, it is simply given by $\mathcal{G}_{L} = L^{M}$, with a factor of $L$ for each magnon. To summarise, this factor reads
\beq\label{NL}
 \N_{L} \rightarrow \prod_{i=1}^{M} \frac{2g}{\sqrt{LE_{i}(E_{i}^2-1)}}\, ,
 \eeq
 and it is of order $O(g^{M})$, for relativistic energies, $E_i = \sqrt{1+(2gp_i)^2} = O(1)$.

\begin{figure}
\begin{center}
\includegraphics[scale=0.35]{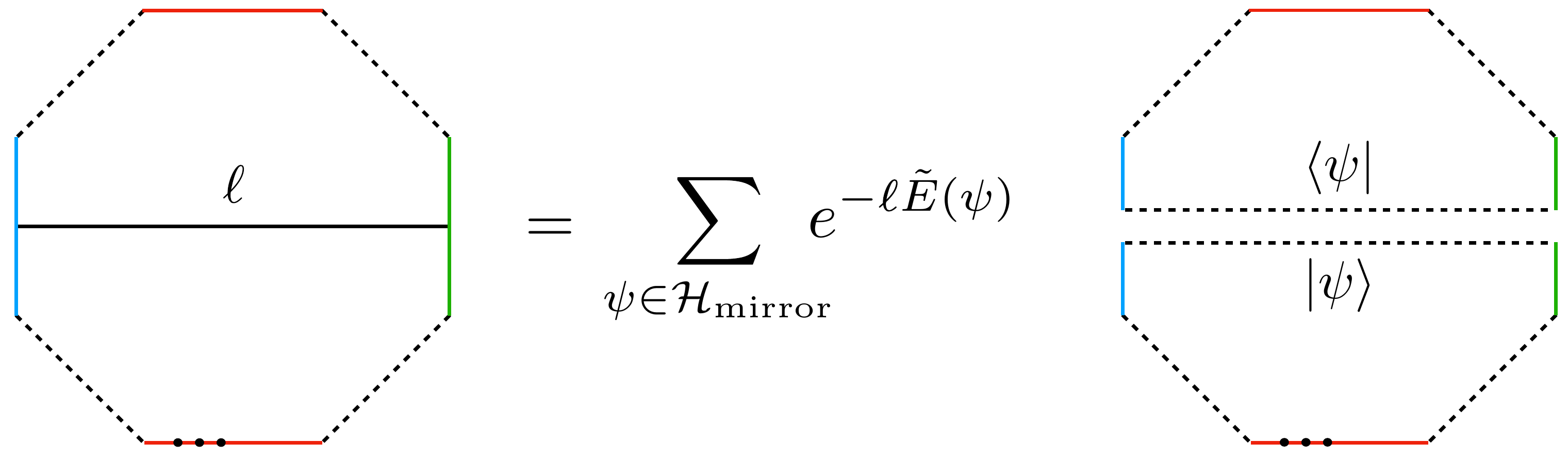}
\end{center}
\caption{The gluing is achieved by inserting a complete basis of mirror magnons on the mirror edge.}\label{sum} 
\end{figure}

Next come the mirror sums. There is no need to detail both of them here since, as we shall argue later on, one follows from the other by analytical continuation. We focus below on the $\B$ sum -- the \textit{bottom or opposing channel} amplitude in the terminology of \cite{Basso:2015eqa} -- which is easier to address and defer its continuation to $\A$ to Section \ref{Sect5}.

Expanding over a basis of mirror states, as depicted in figure \ref{sum}, yields the form factor series for the $\B$ gluing,
\beq
\B(\ell) = 1+ \B_{1}(\ell) + \B_{2}(\ell) + \ldots\, ,
\eeq
where $\B_{n}(\ell)$ is the amplitude for $n$ mirror magnons crossing the bridge of length $\ell = \ell_{B}$. It is described by a $n$-fold sum-integral over the mirror phase space
\beq\label{Bn}
\B_{n} (\ell) = \frac{1}{n!}\sum_{a_{1}, \ldots , a_{n}\geqslant 1} \int \prod_{i=1}^{n}\frac{\tilde{\mu}_{a_{i}}(u_{i})du_{i}}{2\pi}e^{-\ell \E_{a_{i}}(u_{i})} \F_{a_{i}}(u_{i})\T_{a_{i}}(u_{i}) \prod_{i<j}^{n}\tilde{\Delta}_{a_{i} a_{j}}(u_{i}, u_{j})\, .
\eeq
Each mirror magnon carries a rapidity $u$ and a bound-state index $a=1, 2, ...$, labelling its $PSU(2|2)^2$ module with dimension $(4a)^2$. Equivalently, it comes equipped with a pair of  complex conjugated Zhukowski variables,
\beq\label{upm}
x^{[\pm a]} = x(u\pm ia/2) \qquad \rightarrow \qquad \frac{u\pm ia/2}{g} = x^{[\pm a]} + 1/x^{[\pm a]} \, ,
\eeq
which determines its energy and momentum,
\beq\label{fullE}
\E_{a}(u) = \log{(x^{[+a]}x^{[-a]})}\, , \qquad \p_{a}(u) = 2u -\frac{2g}{x^{[+a]}}-\frac{2g}{x^{[-a]}}\, .
\eeq
The latter energy couples in (\ref{Bn}) to the length $\ell$ of the bridge crossed by the particle. It becomes relativistic at strong coupling, for a momentum $\p\sim 1$,
\beq\label{rel}
\E_{a}(\tilde{p}) \sim \frac{\sqrt{\p^2+a^2}}{2g}\, .
\eeq
This low-energy limit corresponds to Zhukowski's close to the unit circle, $x^{[-a]} \rightarrow 1/x^{[+a]}$.

The magnons are weighted in (\ref{Bn}) with a multi-particle measure, which is the mirror image of the one in $\N_{L}$, up to the Gaudin norm. It comprises the individual measure
\beq\label{mua}
\tilde{\mu}_{a}(u)du = \frac{ik(x^{[+a]}, x^{[-a]})du}{(x^{[+a]}-1/x^{[+a]})(x^{[-a]}-1/x^{[-a]})} = \frac{8a^2 g^2 d\p}{(\p^2+a^2)^{2}\sqrt{1+16g^2/(\p^2+a^2)}} \, ,
\eeq 
and the pairwise interaction
\beq\label{Delta}
\tilde{\Delta}_{ab}(u, v) = \prod_{\sigma, \tau = \pm } k(x^{[\sigma a]}, y^{[\tau b]})\, ,
\eeq
where
\beq\label{kxy}
k(x, y) = \frac{x-y}{1-xy}\, .
\eeq
The above interaction is of order $O(g^{0})$ at strong coupling in all the regimes of interest. One also notices that it goes away as soon as one magnon is relativistic,
\beq\label{Dto1}
\lim\limits_{x^{[+a]}x^{[-a]} \rightarrow 1}\tilde{\Delta}_{ab}(u, v) \rightarrow 1\, ,
\eeq
much like its spin-chain analogue, see Eq.~(\ref{hD}). This important property follows from the crossing involution of $k$,
\beq
k(x, y) k(1/x, y) = 1\, ,
\eeq
which can be read out from (\ref{kxy}).

The final ingredient stands for the interaction between a mirror magnon $x^{[\pm a]}$ and the spin-chain magnons at the boundary. It is customarily split into a diagonal part and a matrix part, $\mathcal{F}_{a}$ and $\T_{a}$. While everything before was rather universal, these ones depend on the channel under study. Their general expressions for the bottom channel are given in Appendix \ref{App1}. Below, we discuss them in the BMN regime.

The diagonal part can be expressed in terms of the BES dressing phase \cite{Beisert:2006ez} which is a rather convoluted function for general kinematics. Fortunately, at strong coupling and for state with $\gamma\sim 1$, we only need the AFS phase \cite{Arutyunov:2004vx} which is significantly simpler. Taking all factors into account, we find
\beq\label{above}
\F_{a}(u) = e^{\frac{1}{2}\gamma \E_{a}(u)} = (x^{[+a]}x^{[-a]})^{\frac{1}{2}\gamma}\, ,
\eeq
up to $1/g$ corrections. Its structure is reminiscent of the shock-wave scattering phase $S_{12} \sim e^{i(p_{1}E_{2}-p_{2}E_{1})}$, see e.g.~\cite{Dubovsky:2012sh,Camanho:2014apa}, after performing a double-Wick rotation of one leg to the mirror kinematics. Strikingly, the resulting amplitude grows with the energy as if the Wick rotation has been done in the wrong way.%
\footnote{This large asymptotic behaviour holds at any coupling, as shown in Appendix \ref{App1}.} This is a signature of the bottom channel amplitude and relates to the fact that the mirror magnons are not standing near the physical ones on the chain, that is on an adjacent mirror channel, but lie instead on the edge that is facing it, see figure \ref{sum}.  As a result, in place of a (euclidean) time delay, we find that the mirror magnons exit the bridge earlier than expected, since (\ref{above}) is equivalent to length shift
\beq\label{Fshift}
\ell \rightarrow \beta = \ell- \tfrac{1}{2}\gamma\, .
\eeq
We will see the consequences of this shift later on.

The matrix part $\T_{a}(u)$ is slightly more involved. It contains a sum over the $4a$ diagonal flavours of the mirror magnon and, according to the conjectures in \cite{Basso:2015zoa}, it can be expressed in terms of the diagonal matrix element of the $PSU(2|2)$ transfer matrix. This eigenvalue was worked out in \cite{Beisert:2006qh} for a general Bethe state and can be cast into the form
\beq\label{Ta1}
\mathcal{T}_{a}(u) = \mathcal{T}(x^{[+a]}, x^{[-a]}) = \mathcal{T}^{+} + \mathcal{T}^{-} +  \mathcal{T}^{\, 0}\, ,
\eeq
where each term here is a rational function of $x^{[\pm a]}$. Their general expressions are obtained by continuing the formula in \cite{Beisert:2006qh} to the mirror kinematics and are given in Appendix \ref{App1}. Here, we give $\T$ for a cyclic state with $\gamma = O(1)$ in the $\psu(1,1|2)$ sector at strong coupling. It yields
\beq\label{calTpm}
\mathcal{T}^{\pm} = \pm\frac{i\Q^{\pm}}{g}\sum_{j=1}^{M}\frac{x_{j}}{(x_{j}^2-1)(x^{[\pm a]}x_{j}-1)} \, ,
\eeq
with $\Q^{\pm}$ accounting for the rapidities on the nested level,
\beq\label{calQpm}
\Q^{\pm}  = \prod_{k=1}^{N} \frac{x^{[\mp a]}-y_{k}}{1/x^{[\pm a]}-y_{k}}\, .
\eeq
If not for the $\Q$ factors, these are just generating functions for the higher conserved charges.%
\footnote{The charges are given by $Q_{r} = \sum_{j=1}^{M} x_{j}^{2-r}/(g(x_{j}^2-1))$ and they are generated at both large or small $x^{[\pm a]}$. The expansion starts with the anomalous dimension $\gamma = 2gQ_{2}$ in either case, if the state obeys the zero-momentum condition~(\ref{ZM}), that is, if $Q_{1} = 0$.}
Notice that they are small at strong coupling, $\T^{\pm} = O(1/g)\, ,\forall x^{[\pm a]}$, as expected for a state standing ``close" to the BPS ground state. The last component, $\mathcal{T}^{\, 0}$, is more bulky but plays fortunately a supporting role. It reads
\beq\label{calT0}
\begin{aligned}
\T^{\, 0} = \frac{\Q^{0}}{\Q^{+}\Q^{-}} \T^{+}\T^{-}\, ,
\end{aligned}
\eeq
where
\beq\label{calQ0}
\Q^{0} = \prod_{k=1}^{N}\frac{1}{(1/x^{[+a]}-y_{k})(1-1/x^{[-a]}y_{k})}\sum_{n=0}^{a-2} \Qy(u^{[a-2-2n]})\, .
\eeq
Here, $u^{[m]} = g(x^{[m]}+1/x^{[m]})$ and $\Qy(u) = \prod_{k=1}^{N} (x-y_{k})(1-1/xy_{k})$ is a Baxter polynomial for the auxiliary rapidities. Notice that $\T^{\, 0}$ is quadratic in the charges and $\sim 1/g^2$. It is negligible in the semiclassical regime $\ell \sim g$ but contributes when $\ell = O(1)$, as explained in Section \ref{Sect4}.

\section{The one-particle integral}\label{Sect3}

In this section we analyse the one-particle integral $\B_{1}(\ell)$ at strong coupling. This contribution controls the entire amplitude in the classical domain $\ell \sim g$, after exponentiation of the mirror series. It captures only a bit of the answer for bridge length $\ell \sim 1$, but hints nonetheless at some important simplifications.

\subsection{Classical bridge}

We shall first walk through the classical regime $\ell \sim g \gg 1$. Precisely, we take the strong coupling limit with $l = \ell/(2g)$ kept fixed. Since the length is large, the integrals in (\ref{Bn}) are dominated by the low-energy modes, with momenta $\tilde{p}_i = O(1)$, spins $a_i =O(1)$ and relativistic energies (\ref{rel}). As said earlier, this kinematics corresponds to Zhukowski variables on the unit circle,
\beq\label{x-circ}
x^{[+a]} \rightarrow x\qquad x^{[-a]} \rightarrow 1/x\, , \qquad \forall a\, ,
\eeq
with $|x| = 1$, and, in this variable, the measure (\ref{mua}) reads
\beq\label{overa}
\tilde{\mu}_{a}(u)du \rightarrow \frac{du(x)}{a}\, , \qquad u(x) = g(x+1/x)\, ,
\eeq
while the energy is given by
\beq\label{rel-e}
\E_{a} \rightarrow a\varepsilon(x)/2g\, , \qquad \varepsilon(x) = \frac{2ix}{x^2-1}\, .
\eeq
The measure scale large $ \sim g$ at given $x$, but its scaling is compensated by the transfer matrix $\T = O(1/g)$. Hence, the resulting integrand is of order $O(1)$. The latter takes the same universal form, regardless of the favours of the magnons in the Bethe state. Namely, the flavour dependence drops out, $\Q^{\pm} \rightarrow 1$, when $x^{[\pm a]}$ approach the unit circle, and we get, see~(\ref{Ta1}) and (\ref{calTpm}),%
\beq\label{T-classical}
\T_{a} \rightarrow \T(x) =  \sum_{j=1}^{M} \frac{ix_{j}^2 (x^2-1)}{g(x_{j}^2-1)(xx_{j}-1)(x-x_{j})}\, , \qquad \forall a\, .
\eeq
We can also disregard the diagonal factor, $\F_{a}\rightarrow 1$, since the length shift it produces is subleading here, $\beta \sim \ell$, see Eq.~(\ref{Fshift}).

Moreover, there are no interactions among mirror magnons in this regime, see Eq.~(\ref{Dto1}). As a result, the mirror sum exponentiates,%
\footnote{This is different than for a state with semiclassical energy $\gamma \sim \sqrt{\lambda}$. In that case, $\T = O(1)$ and the pinch singularities in the multi-particle integrals produce sizeable one-particle-like contributions, see \cite{Jiang:2016ulr}.}
\beq
\log{\B(\ell)} \simeq \B_{1} +O(1/g) \, ,
\eeq
with
\beq
\begin{aligned}\label{logBs}
\B_{1}(\ell) &\simeq \int_{C^{+}} \frac{du(x)}{2\pi} \T(x)  \sum_{a=1}^{\infty} \frac{1}{a}e^{-al \varepsilon(x)} \\
& = \sum_{j=1}^{M}\frac{x_{j}^2}{x_{j}^2-1}\int\limits_{\U} \frac{dx}{2\pi i} \frac{(x-1/x)^2}{(x-x_{j})(x x_{j}-1)}\log{(1-e^{-l\varepsilon(x)})}\, ,
\end{aligned}
\eeq
where $C^{+}$ goes from $x=-1$ to $x=+1$ along the unit circle with positive imaginary part. Introducing hyperbolic rapidities,
\beq\label{vareps}
x = \textrm{coth}{\, \tfrac{1}{2}(\theta-\tfrac{i\pi}{2})}\, , \qquad \varepsilon(x) = \cosh{\theta}\, , 
\eeq
and similarly for the Bethe roots, see relations listed after Eq.~(\ref{Eipi}), we obtain the equivalent form
\beq\label{Bd}
\lim\limits_{g\rightarrow \infty}\B(\ell = 2 g \l) = \prod_{j=1}^{M}\d_{l}(\theta_{j})\, ,
\eeq
where $\d_{l}(\theta_{j})$ is given by
\beq\label{logd}
\log{\d_{l}(\theta_{j})} = \int\limits_{-\infty}^{\infty}\frac{d\theta}{2\pi} \frac{f(\theta_{j}+i\frac{\pi}{2})}{f(\theta)\cosh{(\theta-\theta_{j})}} \log{(1-e^{-l\cosh{\theta}})}\, ,
\eeq
with the weight $f(\theta) = \cosh^{2}{\theta}$. Note for later purposes that the coefficient $\d_{l}(\theta)$ obeys a simple functional relation,
\beq\label{crossd}
\d_{l}(\theta_{j}+i\pi)\d_{l}(\theta_{j}) = (1-e^{-il\sinh{\theta_{j}}})\, , \qquad \forall f\, .
\eeq
It relates to the crossing property of the amplitude, see Section \ref{Sect5}.

The factorisation in (\ref{Bd}) is reminiscent of the one observed for the Neumann coefficients of the pp-wave SFT vertex, see \cite{Bajnok:2017mdf,Bajnok:2015hla,Bajnok:2015ftj,Lucietti:2004wy} and references therein. Furthermore, the coefficient $\d_l(\theta)$ in~(\ref{Bd}) and the one denoted $d_{l}(\theta)$ in \cite{Bajnok:2017mdf} which captures the bridge corrections to the Neumann coefficients appear quite similar. The latter also solves the crossing equation (\ref{crossd}) and, as such, can be cast into the integral form (\ref{logd}). Yet it corresponds to a different solution, with a different weight $f$. Namely, in the SFT context, it is the boost-invariant solution $f=1$ that is picked,
\beq
d_{l}(\theta) = b_{l}(\theta)|_{f\rightarrow 1}\, .
\eeq
Despite this difference, there are many common points and as we will see in Section \ref{Sect5} the $b$'s enter the structure constants much like the $d$'s in the Neumann coefficients.

Formula (\ref{Bd}) resums all the mirror corrections $\B_{n} \sim e^{-n l}$ at strong coupling $\forall l \in (0, \infty)$. It shows that $\B$, which begins at $1$ for $l\sim \infty$, grows with decreasing length $l$, all the way to $l=0$ where it blows up. At this end point, it exhibits a power-law behaviour,
\beq\label{small-l}
\B \sim \left(2l\right)^{-\frac{1}{2}\gamma}\, \prod_{j=1}^{M}\cosh{(\tfrac{1}{2}\theta_{j})} \, .
\eeq
This is a rather common short-distance scaling for a form factor series. What is unusual is that it originates from the sum over the bound states and not from large rapidities. Yet another uncommon feature is that the exponent depends on the state at the boundary, through the anomalous dimension $\gamma = \sum_{j} (\cosh{\theta_{j}}-1)$. It can be traced back to the $\theta$ dependence of $f$. It would be absent for a boost-invariant weight $f = 1$, which also brings (\ref{small-l}) for $l \sim 0$ but with $\gamma\rightarrow -1$, no matter the state $\{\theta_j\}$.

To conclude, let us mention the connection with the string theory result \cite{Kazama:2016cfl}. The latter holds for classical Bethe states with energy $\sim g$ dual to classical strings in $AdS_{5}\times S^{5}$. Picking a purely scalar state for simplicity, it predicts that
\beq
\log{C^{\circ\circ\bullet}}|_{\textrm{mirror}} = \mathsf{A}(l) + \textrm{other channels}\, ,
\eeq
where
\beq\label{areaB}
\mathsf{A}(l) = \int_{\U} \frac{du(x)}{2\pi} \{2\textrm{Li}_{2} (e^{-l\varepsilon(x)})-\textrm{Li}_{2} (q_{+}(x)e^{-l\varepsilon(x)}) - \textrm{Li}_{2} (q_{-}(x)e^{-l\varepsilon(x)})\}\, ,
\eeq
with $\textrm{Li}_{2}$ the dilogarithm. In this formulation, all the information about the (zero-momentum) state is encoded in the twists $q_{\pm}(x)$,
\beq\label{qpm}
q_{\pm}(x) =  \exp{\{ \mp \sum_{j=1}^{M}\frac{ix_{j}}{g(x_{j}^2-1)(x^{\pm 1} x_{j}-1)}\}}\, .
\eeq
Since the number of magnons is large for a classical string, $M \sim g \gg 1$, the sums here are of order $O(1)$ and should be read as integrals over the dense support of the roots. Drawing inspiration from the discretisation of the spectral curve introduced in \cite{Arutyunov:2004vx}, we may extrapolate the result to a state with fewer magnons, and energy $\sim 1$, by keeping the exponents as in (\ref{qpm}) and expanding at large $g$. This readily maps $\mathsf{A}(l)$ into (\ref{logBs}), after linearising the twists and differentiating the dilogs. We could also check this reduction, with some more control on the approximation, directly at the level of the hexagon series, which was shown to reproduce $\mathsf{A}(l)$ for classical states in \cite{Jiang:2016ulr}. It links then to the map between the classical and the BMN transfer matrices, $\T(x)_{\textrm{classical}} = 2-\sum_{i=\pm}\exp{\{-\T^{i}(x)_{\textrm{N-BMN}}\}}$, which embodies the exponentiation observed in (\ref{qpm}).

\subsection{Quantum bridge}

We turn to the finite-bridge regime $\ell = O(1)$. Classically, this is the point $l=0$ where the amplitude blows up. We will see here how this singularity is ``resolved" at higher energy, when the non-relativistic corrections, to the dispersion relation notably, are taken into account.

Let us first work out the kinematics. For a finite bridge, the energy should be of order $O(1)$,
\beq\label{disp-rel}
\E_{a}^2 \sim (\p^2+a^2)/4g^2\sim 1/\ell^2\, ,
\eeq
implying that $\p$ or $a$ should be $\sim g$. In either case we observe that the Zhukowski's $x^{[\pm a]}$'s are moving away from the unit circle. These magnons with large individual momenta are the mirror - or AdS - analogues of the spin-chain giant magnons \cite{Hofman:2006xt}. They map to solitonic solutions of the double Wick rotated worldsheet theory and were constructed classically in \cite{Arutyunov:2007tc} at small $a$.%
\footnote{The AdS soliton found in \cite{Arutyunov:2007tc} carries a large momentum $\tilde{p}\sim g$ but no spin. Its bound state analogue should also rotate in AdS when $a\sim g $.}
Their characteristic feature is that they are non-compact: They stretch in AdS and reach the boundary at infinite momenta, as one can see classically using the soliton of \cite{Arutyunov:2007tc}. As such, they can trigger short-distance singularities in the boundary theory, as we shall see later on.

In our case, it appears necessary to have $a \sim g$, as otherwise the measure would be suppressed,
\beq
\tilde{\mu}_{a}(u)du  \sim \frac{a^2 d\p}{g^{2}}\, .
\eeq
see Eq.~(\ref{mua}). This condition was also suggested by the limit $l\rightarrow 0$ of the classical integral (\ref{logd}), which, as said earlier, is dominated by the large $a$'s. In these circumstances, we can approximate the bound-state sum by an integral, $\sum_{a} \rightarrow \int da$, and substitute to the sum over the 1-particle states the 2d integral over the Zhukowski's,
\beq\label{rule}
\sum_{a} \int \frac{\tilde{\mu}_{a}(u) du}{2\pi} \rightarrow ig \int \frac{dx^{[+a]}dx^{[-a]}}{\pi x^{[+a]}x^{[-a]}} k(x^{[+a]}, x^{[-a]})\, ,
\eeq
where we used (\ref{mua}) for the measure and performed a change of variables using (\ref{upm}). Recall that $k(x, y) = (x-y)/(1-xy)$. 

\begin{figure}[!h]
\begin{center}
\includegraphics[width=0.7\textwidth]{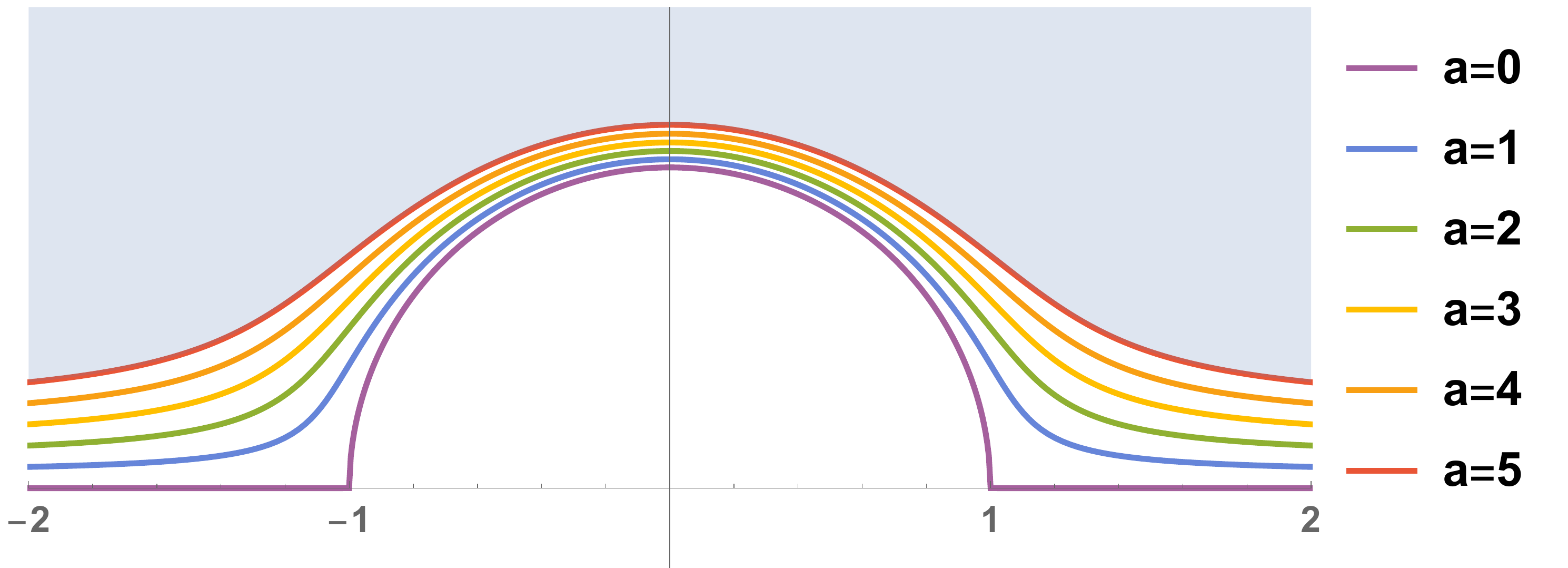}
\end{center}
\caption{Graphs of the Zhukowski variable $x^{[+a]} = x(u+ia/2)$, for real $u$, $a = 1, 2, \ldots\, $ and $g =10$. The short-distance regime is controlled by the continuum formed by this dense semi-infinite set, shown here in blue-grey. Because of a singularity at the boundary, the continuum approximation must stop somewhere over the rainbow, a short distance away from the unit circle.}\label{fig-xplus}
\end{figure}

Combining all the pieces together gives us the one-particle integral for a quantum bridge,
\beq\label{B1s}
\B_{1} = \int \frac{dx^{[+a]}dx^{[-a]}}{\pi (x^{[+a]}x^{[-a]})^{\b+1}} k(x^{[+a]}, x^{[-a]}) \times ig \T (x^{[+a]}, x^{[-a]})\, ,
\eeq
where $\beta = \ell-\tfrac{1}{2}\gamma$ and with $x^{[-a]} = (x^{[+a]})^*$. The domain of integration is the upper half plane for $x^{[+a]}$, minus the unit disk, see figure \ref{fig-xplus}, with the relativistic low-energy modes sitting on the unit circle at the boundary. Note that $g\T, k$, and thus the whole integrand, are of order $O(g^0)$ throughout the entire domain. 

A remarkable property of this integrand is its reflection symmetry under $a\rightarrow -a$. Namely, we see from~(\ref{calTpm}) that 
\beq\label{parity}
\T^{\pm} (x^{[-a]}, x^{[+a]}) = - \T^{\mp} (x^{[+a]}, x^{[-a]})\, .
\eeq
The sum $(\T^{+}+\T^{-})$ is thus antisymmetric and so is the measure, since $k(x, y) = -k(y, x)$. The same can be said about $\T^{\, 0}$, in the regime $a \sim g$, see Eq.~(\ref{calQ02}) below. Hence, since the integrand is a symmetric function of $a$, we can extend the integration domain to the entire plane minus the disk. Introducing polar coordinates,
\beq\label{rs}
x^{[\pm a]} = r\,s^{\pm 1}\, ,
\eeq
with $s = e^{i\phi}$ and $r\in [1, \infty]$, we arrive at
\beq\label{cont}
\B_{1} = \int\limits_{1}^{\infty} \frac{dr}{r^{2\b +1}} \oint\limits_{|s|=1} \frac{ds}{2\pi i s} k(rs, r/s) \times ig \T (rs, r/s)\, ,
\eeq
with the $s$-contour going clockwise along the unit circle.

We will now evaluate this two-fold integral for $\T \rightarrow \T^{+}+\T^{-}$, deferring the integration of $\T^{\, 0}$ to the next section. (The latter component is quadratic in the charges and combines naturally with the higher-magnon integrals.)

Given the parity property (\ref{parity}), we only need to consider the integral for $\T^+$. Focussing on the angular integral, and using (\ref{calTpm}), we get to evaluate
\beq
\mathcal{R}_{1} = \sum_{j=1}^{M} \frac{x_{j}}{x_{j}^2-1} \oint\limits_{|s|=1} \frac{ds}{2\pi i s}\frac{k(rs, r/s)}{1-rs x_{j}} \Q^{+}(rs, r/s)\, .
\eeq
The integrand is a rational function of $s$ with poles at
\beq
\{0, 1/rx_{j}, 1/ry_{k}, \infty\}\, .
\eeq
Recall that all the roots are $>1$ in absolute value; hence, if not for the last one, all the poles in the list above sit inside the unit disk.%
\footnote{Note that one can relax the condition that the $y$'s are outside the disk, as long as they obey the BAEs, since then the associated residues vanish.} So, by Cauchy theorem, the integral follows from its behaviour at infinity,
\beq
\lim\limits_{s\rightarrow \infty}\frac{k(rs, r/s)}{1-rs x_{j}} \Q^{+}(rs, r/s)  = \frac{1}{x_{j}(r^2-1)}\, ,
\eeq
and thus
\beq\label{int-r}
\mathcal{R}_{1} =  \frac{1}{r^2-1} \times \sum_{j=1}^{M}\frac{1}{(x_{j}^2-1)} = \frac{\tfrac{1}{2}\gamma}{r^2-1}\, .
\eeq
Remarkably, all the charges in $\T^{+}$ have been swept away by the angular average, if not for the leading one $\gamma$. We will see shortly that this phenomenon extends to the higher-magnon integrals, which turn out to be proportional to higher powers of $\gamma$ only.

We proceed with the integration over $r$, after doubling (\ref{int-r}) for $\T^{+}+\T^{-}$. Here, we face the problem that the integral does not converge, since $\mathcal{R}_{1}$ has a pole at $r=1$. We regularise it by introducing a lower cut-off $\rm =  1+\epsilon$, with $\epsilon \sim 0$. It gives
\beq\label{B1cut}
\B_{1}|_{\textrm{cut}} =  2\int\limits_{1+\epsilon}^{\infty} \frac{dr}{r^{1+2\b}}\, \mathcal{R}_{1} = -\frac{\gamma}{2}\log{(2\epsilon e^{\gamma_{E}})} -\frac{\gamma}{2}\frac{\partial}{\partial \b}\log{\Gamma(1+\b)}\, ,
\eeq
where $\gamma_{E} = -\log{\Gamma}'(1)$ is the Euler-Mascheroni constant.

The logarithmic divergence at $r=1$ goes back to the $1/a$ scaling of the measure in the classical regime, see Eq.~(\ref{overa}). It indicates that our approximation (\ref{rule}) is not valid in this neighbourhood. Hence, to complete our calculation and get rid of the discrete cut-off dependence in (\ref{B1cut}), we shall reinstate the bound-state sum for the modes close to the disk.

Precisely, we divide the domain into two regions, for the IR and UV modes,
\beq\label{split}
\B_{1} = \B_{1}|_{\textrm{discrete}} +  \B_{1}|_{\textrm{continuum}}\, ,
\eeq
where the continuum part is nothing but the integral (\ref{B1cut}) with $\rm = 1+\epsilon$. It is completed here by the sum over the low-lying bound states, $a = 1, \ldots, \am$, with $\am$ large but finite. To ensure the continuity at the boundary of the sum and the integral, we demand that
\beq\label{matching}
\am = -ig (s-1/s)(\rm-1/\rm) \simeq \frac{4g \epsilon}{\varepsilon(s)}\, ,
\eeq
using the coordinates transformations~(\ref{upm}) and (\ref{rs}). This gluing condition is consistent with our assumptions, as long as $1\ll \am \ll g$ or equivalently $1/g \ll \epsilon \ll 1$. Note that it implies that $\am$ depends on the position $s$ on the unit circle, through the classical energy~(\ref{rel-e}). Other than that, the integrand of the discrete part is as for the classical limit, if not that the bridge is small. Namely, it reads as the first line of~(\ref{logBs}), with $l = 0$,  $x\rightarrow s$ and with a sum up to $\am$.
The sum is readily done
\beq
 \sum_{a = 1}^{\am} \frac{1}{a} \simeq \log{(\am e^{\gamma_{E}})}  = \log{\big[\frac{4g \epsilon e^{\gamma_{E}}}{\varepsilon(s)}\big]}\, ,
\eeq
and so is the integral,
\beq\label{Bdis}
\B_{1}|_{\textrm{discrete}} = \int\limits_{\U} \frac{du(s)}{2\pi} \T(s) \log{\big[\frac{4g \epsilon e^{\gamma_{E}}}{\varepsilon(s)}\big]}  =  \frac{\gamma}{2}\log{(2g\epsilon e^{\gamma_{E}})} + \sum_{j=1}^{M} \log{\cosh{(\tfrac{1}{2}\theta_j})}\, .
\eeq
Hence, as expected, the logarithmic divergences cancel out between (\ref{Bdis}) and (\ref{B1cut}).

The final expression for the one-particle integral is obtained by collecting the finite parts in (\ref{split}). It reads
\beq\label{B1one}
\B_{1} = \log{C}  -\frac{\gamma}{2}\frac{\partial}{\partial \b}\log{\Gamma(1+\beta)}\, ,
\eeq
with $\beta = \ell-\tfrac{1}{2}\gamma$ and with the $\ell$ independent constant
\beq\label{Cst}
C = g^{\frac{1}{2}\gamma} \prod_{j=1}^{M} \sqrt{\tfrac{1}{2}(1+E_{j})} = g^{\frac{1}{2}\gamma} \prod_{j=1}^{M} \cosh{(\tfrac{1}{2}\theta_{j})}\, ,
\eeq
where $E_{j} = \cosh{\theta_{j}}$. The funny scaling of $C$ with $g$ is the imprint left by the singularity at $r=1$. It responds to the logarithmic singularity $\sim \frac{\gamma}{2}\log{(1/l)}$ found classically. This singularity is replaced, as far as the $\ell$ dependence is concerned, by a series of poles,
\beq
\log{\ell} \rightarrow \psi(1+\ell-\tfrac{1}{2}\gamma)\, ,
\eeq
with $\psi$ the digamma function. This is the first step towards the Gamma functions mentioned in the introduction.

Let us mention finally that one can double check (\ref{B1one}) using, instead of a hard cut-off, a modified measure
\beq
\frac{1}{x^{[+a]}x^{[-a]}-1} \rightarrow \frac{1}{(x^{[+a]}x^{[-a]}-1)^{1-\alpha}}\, ,
\eeq
with $\alpha$ a regulator, to be sent to zero at the end of the calculation. The analysis in this scheme is performed in Appendix \ref{reg-app} for completeness.

\section{Summing the multi-particle integrals}\label{Sect4}

In this section we generalise the analysis to the multi-particle exchanges and resum the full form factor series, for a boundary Bethe state in the $\psu(1,1|2)$ sector.

\subsection{Free energy}

For more magnons, it proves convenient to use the Pfaffian formula for the hexagons \cite{Basso:2017khq}, see also \cite{Kostov:2019stn,Kostov:2019auq} for recent discussions. Namely, defining
\beq
z_{i} = x_{i}^{[+a_{i}]}\, , \qquad \bar{z}_{i} = x_{i}^{[-a_{i}]}\, ,
\eeq
for the variables of the $i$-th mirror magnons, we can write the interactions among magnons in the form
\beq\label{DePf}
\prod_{i<j}^{n}\tilde{\Delta}_{a_{i}, a_{j}}(u_{i}, u_{j}) = \prod_{i=1}^{n} \frac{1}{k(z_{i}, \bar{z}_{i})} \times \mathtt{Pf}_{n} \, ,
\eeq
with $k$ as in (\ref{kxy}) and where $\mathtt{Pf}_{n}$ is the Pfaffian of the $2n \times 2n$ antisymmetric matrix with $ij$ element
\beq
k(t_i,t_j)\,,\qquad\quad  t_i= 
\begin{cases}
z_{[i/2]} & i/2 \ \text{non-integer} \\
\bar{z}_{[i/2]}& i/2 \ \text{integer} 
\end{cases}  \qquad i = 1, \cdots, 2n \,.
\eeq
Explicitly, 
\beq \label{eqn-Pfaffian}
\mathtt{Pf}_{n} = \frac{1}{2^n n!} \sum_{\sigma \in S_{2n}} (-1)^{|\sigma|} \prod_{j=1}^n k(t_{\sigma(2j-1)}, t_{\sigma(2j)})\, ,
\eeq
where $S_{2n}$ is the symmetric group of the $2n$ variables $t_{j}$ and $(-1)^{|\sigma|}$ is the signature of the group element $\sigma$.

We should multiply (\ref{eqn-Pfaffian}) with a string of transfer matrices $\prod_{j}\T(z_{j}, \bar{z}_{j})$ and integrate each pair of variables with a measure. Combining them with the product of $1/k$'s coming from the interaction (\ref{DePf}) yields
\beq
\prod_{j=1}^{n}\int \frac{dz_j d \bar{z}_j}{2\pi} \frac{ig \T(z_{j}, \bar{z}_{j})}{(z_j \bar{z}_j)^{1+\b}}\, ,
\eeq
for the individual weights. We can achieve a further important simplification by using permutation symmetry and parity invariance $z_j \leftrightarrow \bar{z}_j$. They allow us to bring each pair of arguments in a given $k$-string in (\ref{eqn-Pfaffian}) to a canonical form of the type $(\bar{z}_{i}, z_{j})$. Hence, below the integral sign, several terms in the Pfaffian can be identified and put together using
\beq
\begin{aligned}
&\mathtt{Pf}_{1} \rightarrow -k(\bar{z}_{1}, z_{1})\, ,\\
&\mathtt{Pf}_{2} \rightarrow k(\bar{z}_{1}, z_{1})k(\bar{z}_{2}, z_{2})-2k(\bar{z}_{1}, z_{2})k(\bar{z}_{2}, z_{1})\, , \\
&\ldots \\
&\mathtt{Pf}_{n} \rightarrow (-1)^n\sum_{\sigma \in S_{n}} 2^{d(\sigma)}(-1)^{|\sigma|} \prod_{i=1}^{n}k(\bar{z}_{1}, z_{\sigma(1)}) \ldots k(\bar{z}_{n}, z_{\sigma(n)}) \, ,
\end{aligned}
\eeq
where $d(\sigma)=\sum_{i}(d_{i}(\sigma)-1)$ with $d_{i}(\sigma)$ the length of the $i$-th cycle in the cycle decomposition of $\sigma$.

Accordingly, $\B$ is akin to a Fredholm determinant and its logarithm is given as a sum over $n$-magnon cycles,
\beq \label{eqn-logB}
\log \B(\ell) = -\sum_{n=1}^\infty \frac{2^{n-1}}{n} \int \prod_{j=1}^n\frac{dz_j d \bar{z}_j}{2\pi} \frac{ig}{(z_j \bar{z}_j)^{1+\b}} (\mathcal{T}_j^{+} + \mathcal{T}_j^{-} + \mathcal{T}^{\, 0}_j) \times C_n\, ,
\eeq
with the cyclic kernel
\beq
C_n =  k(\bar{z}_{1}, z_{2})k(\bar{z}_{2}, z_{3})\ldots k(\bar{z}_{n}, z_{1})\, .
\eeq
The domain of integration is the complex plane minus the unit disk in each variable.

As seen earlier, the 1-magnon integral is problematic at $r = \sqrt{z\bar{z}}= 1$ and necessitates the use of a regulator. The problem is manifest in $C_{1} = k(\bar{z}, z) = r(s-1/s)/(r^2-1)$. It is of no concern when dealing with the component $\T^{\, 0}$, since, as shown below, this one has a zero at $r= 1$. The problem is also absent for the multi-particle integrals in (\ref{eqn-logB}) which do not contain $k$-factors with conjugated arguments. Therefore, all the integrals considered in this section can be taken over the entire domain, with no regulator.

Finally, let us quote the expression for $\T^{\, 0}$ in the regime of interest. As said earlier, this component is naively small at strong coupling, since $\sim 1/g^2$, see Eq.~(\ref{calT0}). However, it comes along with a sum of $\sim a$ terms, see Eq.~(\ref{calQ0}), which enhances the result for $a\sim g$. Replacing the sum in $\Q^{0}$ by an integral makes it clear,
\beq\label{calQ02}
\Q^{\, 0}(z, \bar{z}) = \frac{-i g}{\prod_{k=1}^{N}(1/z-y_{k})(1-1/\bar{z}y_{k})} \int\limits_{\bar{z}+1/\bar{z}}^{z+1/z}dv \prod_{k=1}^{N}(v -(y_{k}+1/y_{k}))\, ,
\eeq 
such that $\T^{\ 0} \sim \T^{\pm} \sim 1/g$. It obviously vanishes at $r =1$, since then $\bar{z} =  1/z$.

\subsection{Telescoping the sum} 
\label{ssub:rest_term}

The $n$-magnon integrand is a polynomial of degree $n$ in the $\mathcal{T}$'s which produces a myriad of terms after opening the brackets in (\ref{eqn-logB}). Fortunately, we will not need to evaluate all of them individually, since, as we will now demonstrate, only the pure powers in $\T^{+}$ or $\T^{-}$ survive in the end, after performing the sum over $n$. The reason is that a cancellation occurs between the ``neutral'' pairs $\T^{+}\T^{-}$ and the $\T^{0}$'s upon integration.

\begin{figure}[!h]
\begin{center}
\includegraphics[scale=0.8]{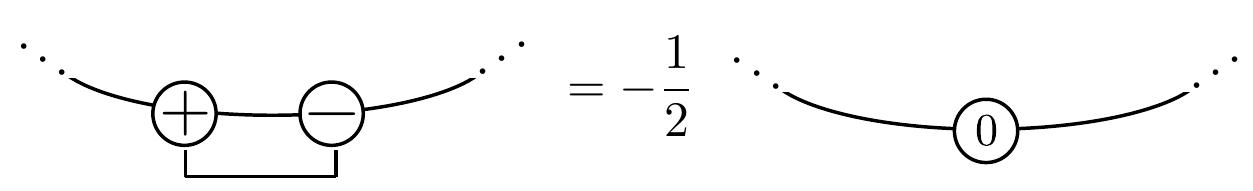}
\end{center}
\caption{Upon integration, a pair $\T^+\T^-$ is equivalent to the insertion of $\T^0$, up to an overall factor, for any given choice of the remaining $\T$'s in the loop.}\label{fig-lemma} 
\end{figure}

The proof goes a follows. First, we shall prove the following identity,
\beq\label{lemma-TT}
\begin{aligned}
&\int \frac{d^2 z_1}{2\pi |z_1|^{2\beta + 2}} \frac{d^2 z_2}{2\pi |z_2|^{2\beta + 2}} k(\bar{z}_\bullet,z_1) \left(ig\mathcal{T}^+_1\right) k(\bar{z}_1,z_{2}) \left(ig\mathcal{T}_2^-\right) k(\bar{z}_2, z_{\circ}) \\
&\qquad \qquad = - \frac{1}{2} \int \frac{d^2z_{2}}{2\pi |z_{2}|^{2\b+2}} k(\bar{z}_\bullet,z_2) \left(ig\mathcal{T}_2^0\right) k(\bar{z}_2, z_{\circ})\, ,
\end{aligned}
\eeq
for fixed $\bar{z}_\bullet, z_\circ$, see figure \ref{fig-lemma}. The equality extends to a length-2 cycle, after identifying the end points, $(z_{\circ}, \bar{z}_{\bullet}) = (z_{2}, \bar{z}_{2})$, and removing the duplicated link $k(\bar{z}_2, z_{\circ})$.

To prove this relation, we introduce polar coordinates, $z_{i} = r_{i} s_{i}, \bar{z}_{i} = r_{i}/s_{i}$, and starts with the integral over $s_{1}$. We evaluate it by picking up the residues outside of the unit disk. There is then only one pole to consider. It comes from $k(\bar{z}_{1},z_{2})$ and sits at $s_{1} = r_{1} z_{2}$. Its residue is obtained using
\beq
-\mathtt{Res} _{s_{1} =\, r_{1} z_{2}} \frac{k(\bar{z}_\bullet, s_{1}r_{1}) k(r_{1}/s_{1},z_{2})k(\bar{z}_{2},z_{\circ})}{s_{1}} = (z_{2}- 1/z_{2}) k(\bar{z}_\bullet, r_{1}^2 z_{2}) k(\bar{z}_{2},z_{\circ})\, ,
\eeq
and evaluating the product $\mathcal{T}^+_{1} \mathcal{T}^-_{2}$ at $(z_{1}, \bar{z}_{1}) = (r_{1}^2z_{2}, 1/z_{2})$,
\beq
\begin{aligned}
ig\mathcal{T}^+ ig\mathcal{T}_j^- & \rightarrow - \sum_{b,c = 1}^{M} \prod_{k=1}^{N} \frac{1/z_{2} - y_{k}}{1/r_{1}^2 z_{2} - y_{k}} \frac{z_{2} - y_{k}}{1/\bar{z}_{2} - y_{k}} \frac{x_b x_c}{(x_b^2-1)(x_c^2-1)(r^2_{1} z_{2} x_b-1)(\bar{z}_{2} x_c -1)} \, .
\end{aligned}
\eeq
Consider next the remaining integrals and rescale $r_{2} \rightarrow r_{2}/r_{1}, s_{2} \rightarrow s_{2}/r_{1}$. It yields
\beq
\int_{1}^{\infty} \frac{dr_{1}dr_{2}}{(r_{1}r_{2})^{2\b+1}} \oint_{|s_{2}|=1} \frac{ds_{2}}{2\pi i s_{2}} \rightarrow \int_{1}^{\infty} \frac{dr_{1}}{r_{1}} \int_{r_{1}}^{\infty} \frac{dr_{2}}{r_{2}^{2\b+1}} \oint_{|s_{2}| = r_{1}} \frac{ds_{2}}{2\pi i s_{2}}\, ,
\eeq
for the measures and contours of integration. The pole structure on the $s_{2}$ plane is such that we can shrink the contour back to $s_{2} =1$ without changing the final result. This is because the only poles enclosed by $|s_{2}| =r_{1}$ are at $s_{2}=0$ and $|s_{2}| = |1/r^2_{1} r_{2} r_{\bullet}| \leqslant 1$.%
\footnote{The latter pole is worrisome when $r_{1} = r_{2} = r_{\bullet} =1$ since it sits along the contour of integration. This situation occurs at the boundary of the domain of integration in $r$ and does not produce any sensible effects.}
We then permute the order of integration for the radial part,
\beq
\int_{1}^{\infty} \frac{dr_{1}}{r_{1}} \int_{r_{1}}^{\infty} \frac{dr_{2}}{r_{2}^{2\b+1}} \oint_{|s_{2}|=1} \frac{ds_{2}}{2\pi i s_{2}} = \int_1^\infty \frac{dr_{2}}{r_{2}^{2\b+1}} \oint_{|s_{2}|=1} \frac{ds_{2}}{2\pi i s_{2}} \int_1^{r_{2}} \frac{dr_{1}}{r_{1}}  \, ,
\eeq
and perform the integration over $r_{1}$. Collecting the terms that are independent of $r_{1}$ after the rescaling yields
\beq
k(\bar{z}_\bullet, z_{2})k(\bar{z}_{2},z_\circ) \prod_{k=1}^{N} \frac{1}{(1/z_{2} - y_{k})(1/\bar{z}_{2}-y_k)} \sum_{b,c} \frac{x_b x_c}{(x_b^2-1)(x_c^2-1)} \frac{1}{(z_{2} x_b-1)(\bar{z}_{2} x_c -1)}\, .
\eeq
The rest gives
\beq
\begin{aligned}
& \int_1^{r_{2}} \frac{dr_{1}}{r_{1}} (r_{1}^2/z_{2}-z_{2}/r_{1}^2) \prod\limits_{k=1}^{N} \left( z_{2}/r_{1}^2 + r_{1}^2/z_{2} -y_k -1/y_k \right)\\
&\,\,\, = -\frac{1}{2} \int_{\bar{z}_{2}+1/\bar{z}_{2}}^{z_{2}+1/z_{2}} dv \prod_{k=1}^{N} \left( v - (y_{k} +1/y_{k})\right)\, ,
\end{aligned}
\eeq
for $v = z_{2}/r_{1}^2 + r_{1}^2/z_{2}$. (This change of variable is a non-self-intersecting path on the complex $v$ plane, implying that we can deform it into a straight line.) This factor is minus the integral part of $\Q^{\, 0}(z_{2}, \bar{z}_{2})$, see Eq.~(\ref{calQ02}), and combining all factors together readily produces the sought-after result. The proof for the length-2 cycle follows similar lines.

\begin{figure}[!h]
\begin{center}
\includegraphics[scale=0.8]{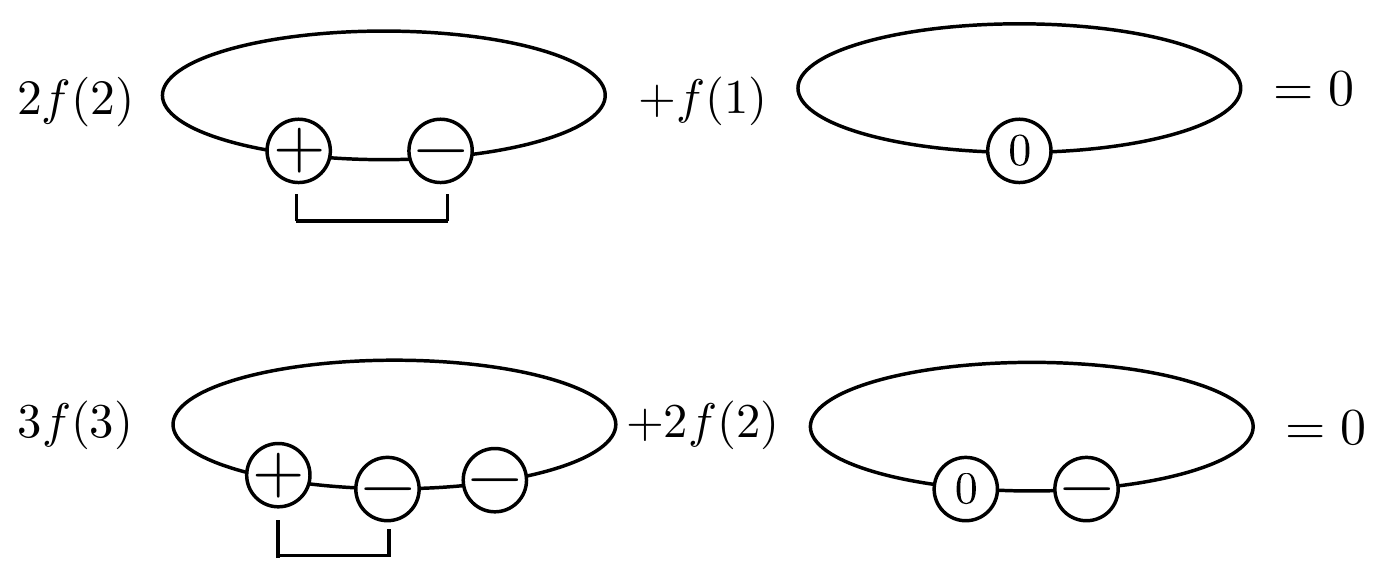}
\end{center}
\caption{Examples of cancellation of loops in the free energy. Here, $f(n) = 2^{n-1}/n$ and the integers are symmetry factors.}\label{fig-cancel} 
\end{figure}

We will now show that relation (\ref{lemma-TT}) leads to the telescoping of the sum (\ref{eqn-logB}) and to the cancellation of all the terms that are not pure powers of $\T^{+}$ or $\T^{-}$. Let then $\mathcal{H}$ be the Hilbert space spanned by the basis $\{| z\rangle: z \in \mathbb{C}, |z|^2>1\}$ with delta-function normalised elements,
\beq\label{resolution}
\langle z | z'\rangle = 2\pi \delta^{(2)} (z-z') \qquad \Rightarrow \qquad \mathds{1}= \int \frac{d^2 z}{2\pi} |z \rangle \langle z |\, ,
\eeq
and let us view the $\T$'s as the matrix elements of certain linear operators $\mathbb{T}$ on $\mathcal{H}$, by defining%
\beq
\langle z_i| \, \mathbb{T}^{\pm, 0} | z_{j} \rangle  =  \frac{ig\mathcal{T}^{\pm, 0}_{i}}{|z_{i}|^{2\b+2}} \times k(\bar{z}_i, z_j) =  \frac{ig\mathcal{T}^{\pm, 0}_{i}}{|z_{i}|^{2\b+2}} \frac{\bar{z}_{i}-z_{j}}{1-\bar{z}_{i}z_{j}}\, .
\eeq
(So defined, the operators are not Hermitian, but we will not need that property here.)
With their help, we can write the free energy as
\beq\label{trlog1}
\begin{aligned}
\log \B(\ell) = \frac{1}{2} \mathtt{tr}_{\mathcal{H}} \log \, (\mathds{1} - 2(\mathbb{T}^+ + \mathbb{T}^- + \mathbb{T}^{\, 0}) ) \, ,
\end{aligned} 
\eeq
where the $\log$ is defined as a power series and the trace $\mathtt{tr}_{\mathcal{H}}$ is taken over $\mathcal{H}$ using (\ref{resolution}). Equation (\ref{lemma-TT}) translates into%
\footnote{The identity holds inside the trace, as it entails contour manipulations and change in the order of integrations.}
\beq
\mathtt{tr}_{\mathcal{H}}\, (\cdots \, \mathbb{T}^{\, 0} \cdots ) = -2\, \mathtt{tr}_{\mathcal{H}} \, (\cdots\, \mathbb{T}^+ \mathbb{T}^- \cdots)
\eeq
and therefore we conclude that
\beq
\log{\B(\ell)} = \frac{1}{2} \mathtt{tr}_{\mathcal{H}} \log \, ((\mathds{1} - 2\mathbb{T}^+) (\mathds{1} - 2\mathbb{T}^-) ) = \mathtt{tr}_{\mathcal{H}} \log  \, (\mathds{1} - 2\mathbb{T}^+)\, ,
\eeq
using, in the last step, that plus and minus terms are related by parity $z\leftrightarrow \bar{z}$ and contribute equally. A cartoon of the cancellation is given in figure \ref{fig-cancel} for a few examples.

\subsection{Evaluating the sum} 
\label{subsub:allplusesterm}

In the end, we are left with the much simpler problem of evaluating the sum over the homogeneous cycles,
\beq
\log{\B(\ell)} =  -\sum_{n\geqslant 1}\frac{2^n}{n}\int C_n\,  ig\mathcal{T}_1^+ ig\mathcal{T}_2^+ \cdots ig\mathcal{T}_n^+\, .
\eeq
As before, we will first carry out the integrals over the angular variables,
\beq
\mathcal{R}_{n} = (-1)^n\sum_{b_1,b_2,\cdots , b_n} \prod_{j=1}^{n} \frac{x_{b_j}}{x_{b_j}^2 -1} \oint\limits_{|\z_1| = 1} \cdots \oint\limits_{|\z_n| = 1} \prod_{j=1}^{n} \frac{d\z_j}{2\pi i \z_j}  \frac{k(\bar{z}_{j}, z_{j+1})}{z_j x_{b_j} -1} \underbrace{\prod_{k=1}^{N}  \frac{\bar{z}_j - y_k}{1/z_j -y_k}}_{= \mathcal{Q}_j^+}\, ,
\eeq
where $z_{n+1} = z_{1}$. We can integrate the variables recursively, from $s_1$ to $s_{n-1}$, by picking up the residues of the poles outside the unit disk at each step. The latter poles only come from the string of $k$-factors,
\beq
1 - \bar{z}_{j} z_{j+1} = 0 \qquad \Rightarrow \qquad s_{j} = r_{j}z_{j+1}\, .
\eeq
Their contributions are read off using an anticlockwise contour and follow from
\beq\label{Resj}
-\mathtt{Res}_{s_{j} = \, r_{j}z_{j+1}} \left[\frac{\Q^{+}_{j}k(\bar{z}_j, z_{j+1})}{s_{j} (x_{b_{j}}z_{j}-1)}\right] = \Q^{+}(z_{j}, \bar{z}_{j})\times \frac{(z_{j+1} - 1/z_{j+1})}{z_{j+1} r^{2}_{j} x_{b_{j}}-1}\, ,
\eeq
with $\Q^{+}$ evaluated at
\beq\label{rec}
(z_{j}, \bar{z}_{j})\rightarrow (r_{j}^2 z_{j+1}, 1/z_{j+1})\, .
\eeq
We can prove recursively that these simple poles are the only ones that contribute. For example, looking at step 1 first, the relevant component of the integrand is
\beq
\frac{k(\bar{z}_{n}, z_{1})\Q_{1}^{+}k(\bar{z}_{1}, z_{2})}{s_{1}(z_{1}x_{b_{1}}-1)}\, .
\eeq
It has poles at $s_{1} = \{r_{1} z_{2}, 0, 1/r_{1}\bar{z}_{n}, 1/r_{1}x_{b_{1}}, 1/r_{1}y_{k}\}$, which all sit inside the unit disk, except for the first one.%
\footnote{We are using here that $r_{j}, x_{b_{j}}, y_{k} \geqslant 1$ and disregard the exceptional situations where the poles are on the circle.} Note that there is no pole at infinity, since the above factor is of order $O(1/s^2_{1})$ when $s_{1} \sim \infty$. Assuming a similar pole structure is found at step $j$, we can prove that no new poles poles are generated for the $s_{j+1}$ integral. To this end, we simply notice that applying the recursion rule (\ref{rec}) to the RHS of (\ref{Resj}) only introduces poles in the unit $j+1$-th disk. (Iterating (\ref{rec}) also shows that this is so for any $\z_k, \forall k \geq j$ as well.) Furthermore, the large $s$ behaviour stays the same and $\mathcal{Q}_{j}^+ = 1 + O(1/s_j) = 1 + O(1/s_k), \forall k \geq j$, while
\beq
\lim\limits_{s_{j+1}\rightarrow \infty} \frac{z_{j+1} - 1/z_{j+1}}{(z_{j+1}r_{j}^2 x_{b_j} -1)} = 1/r_{j}^2 x_{b_{j}}\, ,
\eeq
and similarly for all $k \geq j$ after iterating with (\ref{rec}).

This is justifying our assumption about the pole structure, leaving us with the $\z_n$ integral
\beq
\mathcal{R}_{n} = (-1)^{n}\sum_{b_1,b_2,\cdots , b_n} \prod_{j=1}^{n} \frac{x_{b_j}}{x_{b_j}^2 -1} \oint\limits_{|\z_n| =1} \frac{d\z_n}{2\pi i \z_n} \frac{k(\bar{z}_{n}, z_1)}{z_{n} x_{b_{n}}-1} \Q^{+}_{n} \prod_{j=1}^{n-1} \Q^{+}_j \frac{(z_{j+1} - 1/z_{j+1})}{z_jx_{b_j} -1}\, ,
\eeq
where, for $j< n$,
\beq
z_j = z_{n}\prod_{k=j}^{n-1} r_k^2, \quad \bar{z}_j = \frac{1}{z_{n}}\prod_{k=j+1}^{n-1} r_k^2\, .
\eeq
The only pole for this integral is at $\z_n=\infty$. Since in this limit, $\Q^{+}_{k} \rightarrow 1, \forall k$ and
\beq
\frac{k(\bar{z}_n,z_1)}{z_{n} x_{b_{n}}-1}  \prod_{j=1}^{n-1} \frac{(z_{j+1} - 1/z_{j+1})}{z_jx_{b_j} -1} \rightarrow \frac{1}{\prod_{j=1}^{n}x_{b_{j}}(\prod_{j=1}^{n} r_j^2-1)}\, ,
\eeq
we conclude that the residue at $s_{n} = \infty$ gives
\beq
\mathcal{R}_{n} =  \frac{(-\tfrac{1}{2}\gamma)^n}{\prod_{j=1}^{n} r_j^2-1}\, .
\eeq
Remarkably, the angular average is proportional to the $n$-th power of the anomalous dimension. For $n=1$, it of course reproduces expression (\ref{int-r}).

We should then dress this result with the radial weight $(r_{1}\ldots r_{n})^{-1-2\b}$ and integrate each variable from $1$ to $\infty$. The radial integration is immediately performed for $n >1$, using
\beq
\begin{aligned}\label{radials}
\int\limits_{1}^{\infty} \frac{dr_1\ldots dr_{n}}{(r_1\ldots r_{n})^{2\b+1}}\times \frac{1}{\prod_{j=1}^{n} r_j^2-1} &= \sum_{k=1}^\infty\bigg[\int\limits_{1}^{\infty} \frac{dr}{r^{2\b+2k+1}}\bigg]^{n} \\
= \frac{1}{2^n} \sum_{k=1}^{\infty} \frac{1}{(\b+k)^n} &= \frac{(-1)^{n}}{2^n (n-1)!}\frac{\partial^{n}}{\partial \beta^n}\log{\Gamma(1+\b)} \, .
\end{aligned}
\eeq
Here, we expanded the geometric series and exchanged the order of summation and integration, which is justified since both the sums and the integrals converge absolutely for $r>1$. For $n=1$ the integral is singular at $r =1$ and we are back to the discussion in Section \ref{Sect3}.

Putting all factors together we arrive at
\beq
\log{\B(\ell)} = \log{C} - \sum_{n=1}^{\infty} \frac{1}{n!}(\tfrac{1}{2}\gamma)^{n} \frac{\partial^{n}}{\partial \beta^n} \log{\Gamma(1+\beta)}\, ,
\eeq
that is,
\beq\label{oct1}
\B(\ell) = C\,  \frac{\Gamma(1+\b)}{\Gamma(1+\b+\tfrac{1}{2}\gamma)} = C\,  \frac{\Gamma(1+\ell -\frac{1}{2}\gamma)}{\Gamma(1+\ell)}\, .
\eeq
It features two Gamma functions, which relate through their arguments to the $\mathfrak{so}(4,2)$ and $\mathfrak{su}(4)$ quantum numbers of the operators, in such a way that $\B(\ell) \rightarrow 1$ in the BPS limit $\gamma \rightarrow 0$.

The prefactor $C$ stems from the regularisation of the logarithmic divergence, which only affects the $n=1$ integral in the free energy. It was determined in Section \ref{Sect3}, see Eq.~(\ref{Cst}), and it is the only piece that depends on more than just the anomalous dimension $\gamma$. Other than that, the continuum approximation works as a mini-superspace approximation by projecting on the global quantum numbers.

Equation (\ref{oct1}) is our final expression for $\B(\ell)$ when $\ell = O(1)$. Taking $\ell$ large in (\ref{oct1}) leads to a power law,
\beq
C\frac{\Gamma(1+\ell -\frac{1}{2}\gamma)}{\Gamma(1+\ell)} \sim C \ell^{-\frac{1}{2}\gamma}\, ,
\eeq
which agrees with the small-$l$ classical scaling (\ref{small-l}). This is not so surprising since the one-particle integral is the only one that survives in the exponent when $\beta\rightarrow \infty$. Barring an order of limit issue, this matching suggests that formula (\ref{oct1}) ``resums'' the leading singularities at $l \rightarrow 0$ of the semiclassical expansion.

\subsection{Changing the grading}

Formula (\ref{oct1}) applies to any $\psu(1,1|2)$ primary with non-zero fermionic roots in the $\su(2)$ grading,
\beq
y_{i}\neq 0, \,\,\,\, \forall i  =1, \ldots , N\, .
\eeq
This condition was used implicitly to show that $\Q^{\pm} \rightarrow 1$ when $x^{[\mp a]}, 1/x^{[\pm a]}\rightarrow 0$ with $x^{[+ a]}x^{[-a]} = r^2$ fixed. Adding a root $y = 0$ is a straightforward operation; it is equivalent to shifting the length,
\beq\label{shift}
y = 0 \qquad \leftrightarrow \qquad  \ell \rightarrow \ell-1\, ,
\eeq
in agreement with a general property of the transfer matrix, see Appendix \ref{App1},
\beq\label{y0}
\T_{a}(u) |_{\{y \rightarrow 0, y_{i}\}} = x^{[+a]}x^{[-a]}\,  \T_{a}(u)|_{\{y_{i}\}}\, .
\eeq
The root $y = 0$ is special in that it is associated to the symmetry transformation \cite{Beisert:2005fw} mapping a primary from the $\su(2)$ to the $\sl(2)$ grading, denoted as $\eta = +1, -1$, respectively, in the following.%
\footnote{A root at infinity should also be added, in principle, but it plays no role in the discussion.}
The general formula that covers both gradings is
\beq\label{Blast}
\B(\ell) = C\,  \frac{\Gamma(\ell + \frac{1+\eta}{2}-\tfrac{1}{2}\gamma)}{ \Gamma(\ell + \frac{1+\eta}{2})}\, .
\eeq
Note that primaries related by $y = 0$ descend from the same superconformal primary, though through different paths,%
\footnote{Schematically, $\mathcal{O}_{\eta = +1} = Q_{1}^{\,\, 1}Q_{2}^{\,\, 1}\bar{Q}_{\dot{1}4}\bar{Q}_{\dot{2}4} \cdot \mathcal{O}_{\textrm{bottom}}$ and $\mathcal{O}_{\eta = -1} = Q^{\,\,1}_{1}Q^{\,\, 2}_{1}\bar{Q}_{\dot{1}3}\bar{Q}_{\dot{1}4} \cdot \mathcal{O}_{\textrm{bottom}}$ with $Q_{\alpha}^{\,\, A}, \bar{Q}_{\dot{\alpha}B}$ the supercharges and with $\mathcal{O}_{\textrm{bottom}}$ a superconformal primary.}
and they must share the same structure constant by diagonal $PSU(2|2)$ symmetry. We can verify it by using the conversion rules given in table \ref{primaries}. One reads that going from $\eta = +1$ to $\eta = -1$ amounts to replacing two scalars $\Y$ by two derivatives $\D$, implying a loss of two units of length. Since $\ell = \ell_{B} = \frac{1}{2}(L_{2}+L_{1}-L)$ and since $\B$ does not depend on the spins, we verify that it is equivalent to (\ref{shift}).

\begin{table}
\begin{center}
\begin{tabular}{|p{2cm}||p{1.5cm}|p{2cm}|p{1.5cm}|p{3cm}|} 
\hline
Primary & Length & Dimension & Spin & (R, Y) charges \\
\hline
bottom & $L$ & $\Delta-2$ & $S-2$ & $(J-2, Y)$ \\ 
$\eta = -1$ & $\color{red}{L}$ & $\color{red}{\Delta}$ & $\color{red}{S}$ & $\color{red}{(J, Y)}$ \\ 
$\eta=+1$ & $L+2$ & $\Delta$ & $S-2$ & $(J, Y+2)$ \\
top & $L+2$ & $\Delta+2$ & $S-2$ & $(J+2, Y)$ \\ 
\hline
\end{tabular}
\smallskip
\caption{List of bosonic primaries used in this paper. They all lie on the diagonal of a super-multiplet and share the same BMN energy = $\sum_{j=1}^{M}E_{j}$, anomalous dimension $\gamma$ and magnon number $M$. We choose as a reference point the quantum numbers of the primary in the $\sl(2)$ grading $\eta=-1$. The bottom operator is the superconformal primary, i.e.~the operator with the lowest dimension in the supermultiplet. The top representative refers here to the primary with maximal R-charge. (Note that neither the bottom nor the top component live inside the $\psu(1,1|2)$ sector.) The $\su(4)$ Dynkin labels $[q, p, q] $ are given by $q = $ Y-charge and $p + q =$ R-charge.}\label{primaries}
\end{center}
\end{table}

\section{HHL structure constants}\label{Sect5}

In this section, we complete the analysis and obtain formulae for the structure constants of interest. We then discuss their main properties and argue that wrapping corrections should be negligible in the HHL regime $L_{2}\ll \ell_{C}\sim g$. 

\subsection{Crossing}

\begin{figure}
\begin{center}
\includegraphics[scale=0.35]{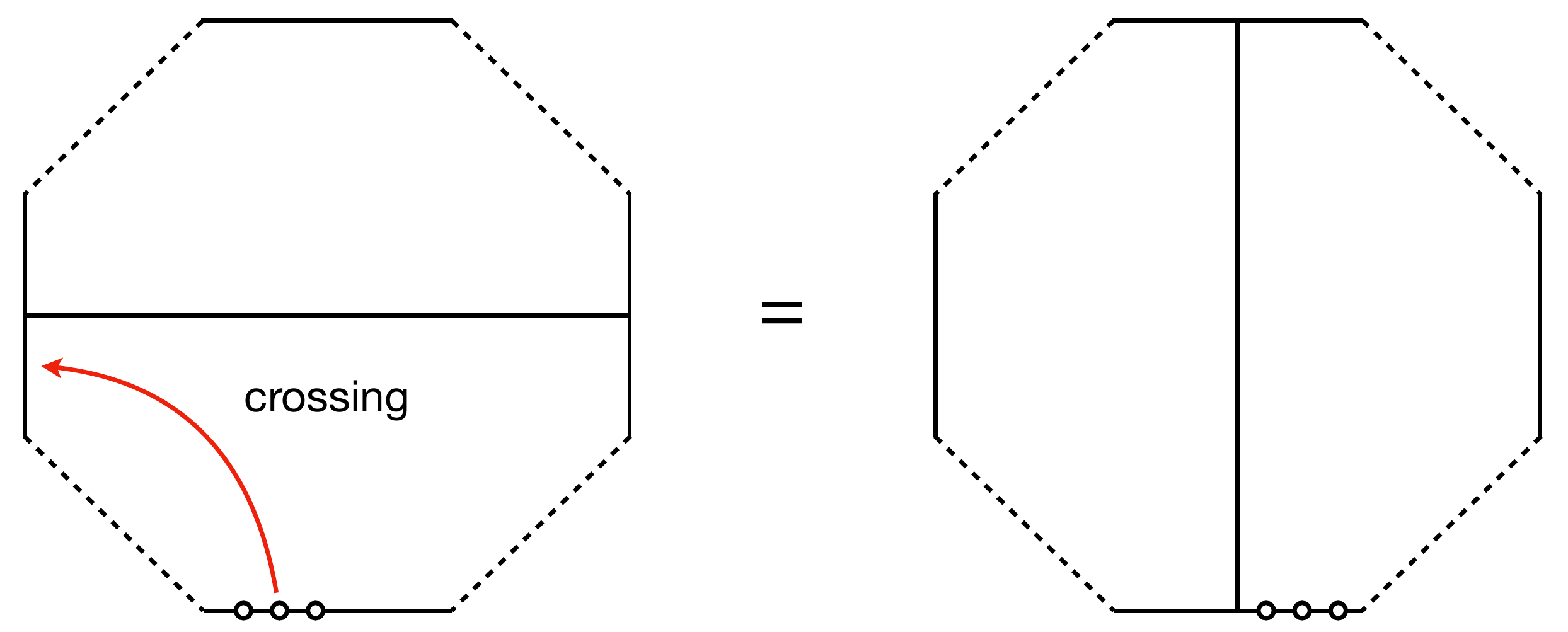}
\end{center}
\caption{Crossing the magnons from one edge to another maps the amplitudes $\B$ and $\A$.}\label{cross} 
\end{figure}

It remains to determine the amplitude $\A$. This one admits a form factor series, much like $\B$, which is spelled out in \cite{Basso:2015eqa}. However, it appears technically harder to analyse it by following the same lines as used for $\B$. The reason is that it involves a crossed version of the transfer matrix, i.e.,
\beq
\T(1/x^{[+a]}, 1/x^{[-a]})\, ,
\eeq
which displays singularities for $x^{[\pm a]} = x_{j}$, as one can see by flipping the Zhukowski's in Eqs.~(\ref{calTpm}). These singularities are harmless for the semiclassical study, which leads straightforwardly to the result quoted below, see Eq.~(\ref{Aclass}). But it makes the analysis much arduous for a quantum bridge, which appears very sensitive to the singularities of $\T$. Fortunately, we can get around it by crossing the real magnons in the state, using
\beq
E_j \rightarrow - E_j\, ,
\eeq
instead of flipping the mirror particles. Geometrically, it corresponds  to transporting the state along the contour of the octagon and maps $\B$ into $\A$, as shown in figure \ref{cross}.%
\footnote{There are a few complications there as well. Firstly, since we are working with scattering eigenstates, we should pay attention to the crossing of the auxiliary roots. However, given that these ones drop out in the final result, we believe we can ignore them here. More importantly, one cannot cross all the magnons along the same path and at the same time satisfy the zero-momentum condition. The latter requirement imposes that magnons rotate in opposite directions. We can get around this problem without relaxing $P=0$ by assuming that the magnons were split from the outset and laid on different edges, at the bottom and top of the octagon in the right panel of figure \ref{cross}. The amplitude $\B$ does not depend on how we partition the state, see \cite{Basso:2015zoa}. The magnons can then be crossed towards the same edge by following different directions, as needed to balance their momenta.}

Let us illustrate the operation for a classical bridge $l = \ell/2g = O(1)$. We then cross all the magnons using
\beq
\theta_{j} \rightarrow \theta_{j} +i\pi\, ,
\eeq 
which immediately yields, using the functional relation (\ref{crossd}),
\beq\label{Aclass}
\lim\limits_{g\rightarrow \infty} \A(\ell  = 2g \l) = \prod_{j=1}^{M}d_{l}(\theta_{j}+i\pi) = \prod_{j = 1}^{M}(1-e^{-i\l\sinh{\theta_{j}}}) \prod_{j=1}^{M} \frac{1}{d_{l}(\theta_{j})}\, .
\eeq
The first factor in the RHS is the asymptotic part of the amplitude, which dominates when $\l \rightarrow \infty$. It is generally written as a sum over $2^M$ partitions of the Bethe roots on the two sides of the bridge \cite{Escobedo:2010xs,Basso:2015zoa} weighted by the hexagon form factors $h$. The sum factorises here because $h\rightarrow 1$ at strong coupling. The next factor falls off exponentially at large $l$,
\beq
d_{l}(\theta) = 1+ O(e^{-\l})\, .
\eeq
It accounts for the mirror magnons crossing the bridge $A$.%
\footnote{In the opposite limit, for $\l\rightarrow 0$, $\A$ has a power-law behaviour, much like $\B$. However, the amplitude is becoming small there, $\mathcal{A}(\ell) \sim l^{M+\frac{1}{2}\gamma} \prod_{j=1}^{M} \sqrt{\tfrac{1}{2}(1-E_{j})}$, unlike $\B$.}

\begin{figure}
\begin{center}
\includegraphics[scale=0.35]{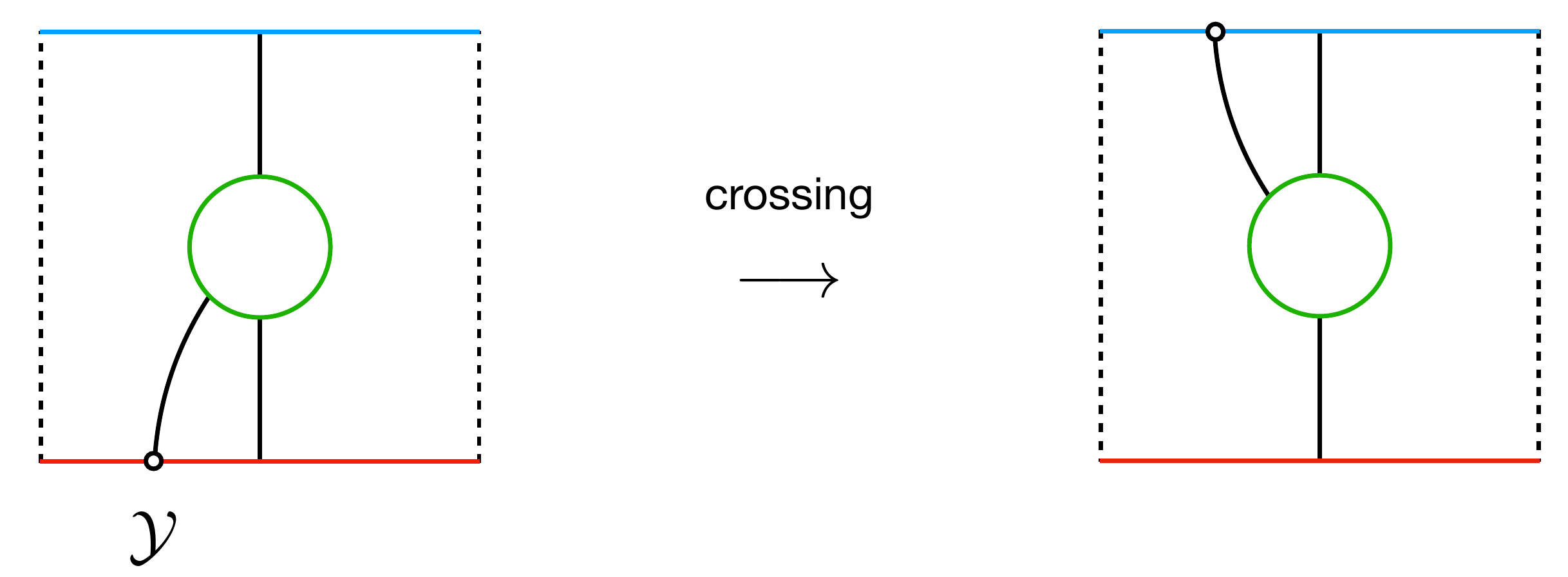}
\end{center}
\caption{Scalar's swing under crossing. The crossing of a scalar field produces jumps in the bridge lengths, which we can interpret as saying that the field is moving along with its propagator, as shown here.}\label{swing} 
\end{figure}

We must proceed more carefully when performing this operation at finite bridge for scalars (or more precisely for a state with $Y\neq 0$). These excitations are known to induce jumps in the spin-chain lengths under crossing. The right thing to do is to cross in the string frame, that is, at fixed R charge, see \cite{Arutyunov:2009ga} for a review. In this frame, the $\Y$'s do not cause any problem; they are as ``lengthless" as the $\D$'s. Replacing the lengths by the R-charges in (\ref{bridges}) gives the splitting lengths,
\beq\label{WSbridges}
j_{A, B} =  \ell_{A, B} \mp \frac{1}{2}Y\, ,
\eeq
where we used $J_{1,2} = L_{1,2}$ and $J = L-Y$. Hence, the correct crossing map is taken at fixed $j$ and reads
\beq\label{cf}
\A(\ell) = \B_{E\rightarrow -E}(\ell-Y) \, .
\eeq
It agrees with the classical transformation, when $Y \ll \ell\sim g$. Notice that we can interpret this shift pictorially as saying that the $\Y$'s carry propagators with them upon crossing, as shown in figure \ref{swing}. Applying this recipe to (\ref{Blast}) for $\eta = -1$ yields the sought-after amplitude,
\beq\label{Aq}
\A(\ell) = g^{-M-\frac{1}{2}\gamma}\prod_{j=1}^{M}\sqrt{\tfrac{1}{2}(1-E_{j})}\, \, \frac{\Gamma(\ell+S+\tfrac{1}{2}\gamma)}{\Gamma(\ell-Y)}\, ,
\eeq
where $S = M-Y$ is the Lorentz spin of the operator. Note that the ratio of Gamma functions no longer goes to $1$ when $\gamma\rightarrow 0$, because the anomalous dimension picks a canonical part under crossing,
\beq\label{gammacross}
\gamma \rightarrow -2M -\gamma\, .
\eeq

We are now equipped to write down the structure constants. Several expressions can be obtained by scaling independently the lengths of the two bridges.%
\footnote{One could also consider processes with excitations on the two heavy operators by crossing part of the state to the other channel.}
We focus here on $\ell_{A, B} = O(1)$. Using expressions~(\ref{Aq}) and (\ref{Blast}) for $\A$ and $\B$ in (\ref{factorise}), we get
\beq\label{get}
C^{\circ\circ\bullet}/C^{\circ\circ\circ} = \M\times \frac{\Gamma(\ell_{B} -\tfrac{1}{2}\gamma)\Gamma(\ell_{A}+S+\tfrac{1}{2}\gamma)}{\Gamma(\ell_{B})\Gamma(\ell_{A}-Y)}\, ,
\eeq
for a primary in the grading $\eta = -1$.%
\footnote{The expression for the $\eta = +1$ primary is obtained by replacing $\ell_{A, B} \rightarrow \ell_{A, B} +1$ everywhere.} It reduces to the form given in (\ref{intro}) when $Y = 0$. Interestingly, the funny powers of $g$ in $\A$ and $\B$ disappear in the product $\A\B$. The remaining prefactors assemble such as to cancel part of the denominator in the norm $\mathcal{N}_{L}$, see Eq.~(\ref{NL}), leaving just
\beq
\M = \prod_{j=1}^{M}(LE_{i})^{-1/2}\, ,
\eeq
up to an overall phase. Hence, if not for the Gamma functions, the magnons are produced independently of each other, with a constant weight (up to the relativistic measure).

Reinstating the quantum numbers of the operators makes the symmetry between the AdS numerators and the sphere denominators more manifest. It yields
\beq
C^{\circ\circ\bullet}/C^{\circ\circ\circ} = \M\times \frac{\Gamma[\frac{1}{2}(\Delta_{2}+\Delta_{1}-\Delta+S)]\Gamma[\frac{1}{2}(\Delta_{2}-\Delta_{1}+\Delta+S)]}{\Gamma[\frac{1}{2}(J_{2}+J_{1}-J-Y)]\Gamma[\tfrac{1}{2}(J_{2}-J_{1}+J-Y)]}\, ,
\eeq
where $\Delta_{i}  = J_{i} = L_{i}$ for the chiral primary operators and with $J = L-Y, \Delta = L+S+\gamma$ for the unprotected operator. The formula could also be written in terms of the weights of other representatives in the supermultiplet using table \ref{primaries}, but at the cost of disgraceful shifts in the arguments of the Gamma functions. The $\eta = -1$ primary appears as the nicest choice in this respect.

\subsection{Poles and zeros}

Formula (\ref{get}) displays a simple pole when
\beq\label{mixing}
\beta = \ell_{B}-\tfrac{1}{2}\gamma  =  -n\, , 
\eeq
for integer $n$. As explained in detail in \cite{Korchemsky:2015cyx}, these poles relate to the mixing between single- and double-trace operators at order $1/N_{c}$. The double-trace operators that are relevant here are those overlapping with the two chiral primaries in the structure constants at order $N_{c}^0$. They are local products of descendants of $\mathcal{O}_{1}$ and $\mathcal{O}_{2}$ in the same Lorentz and R-symmetry representations as $\mathcal{O}_{\gamma}$ and read, schematically, 
\beq\label{mixing}
\mathcal{O}_{DT} \sim \textrm{tr}\, (\Z^{L_{1}})\, \Box^{n} \D^{S}\textrm{tr}\, (\Z^{L_{2}-\ell_{B}-Y}\Y^{Y}\bar{\Z}^{\ell_{B}}) + \ldots\, ,
\eeq
with $\Box = \D\bar{\D}+\ldots\,$ the Laplacian and with the dots indicating the need to mix the fields properly. The mixing occurs when the scaling dimensions of $\mathcal{O}_{\gamma}$ and $\mathcal{O}_{DT}$ are matching, that is precisely when (\ref{mixing}) is satisfied.

\begin{figure}
\begin{center}
\includegraphics[scale=0.35]{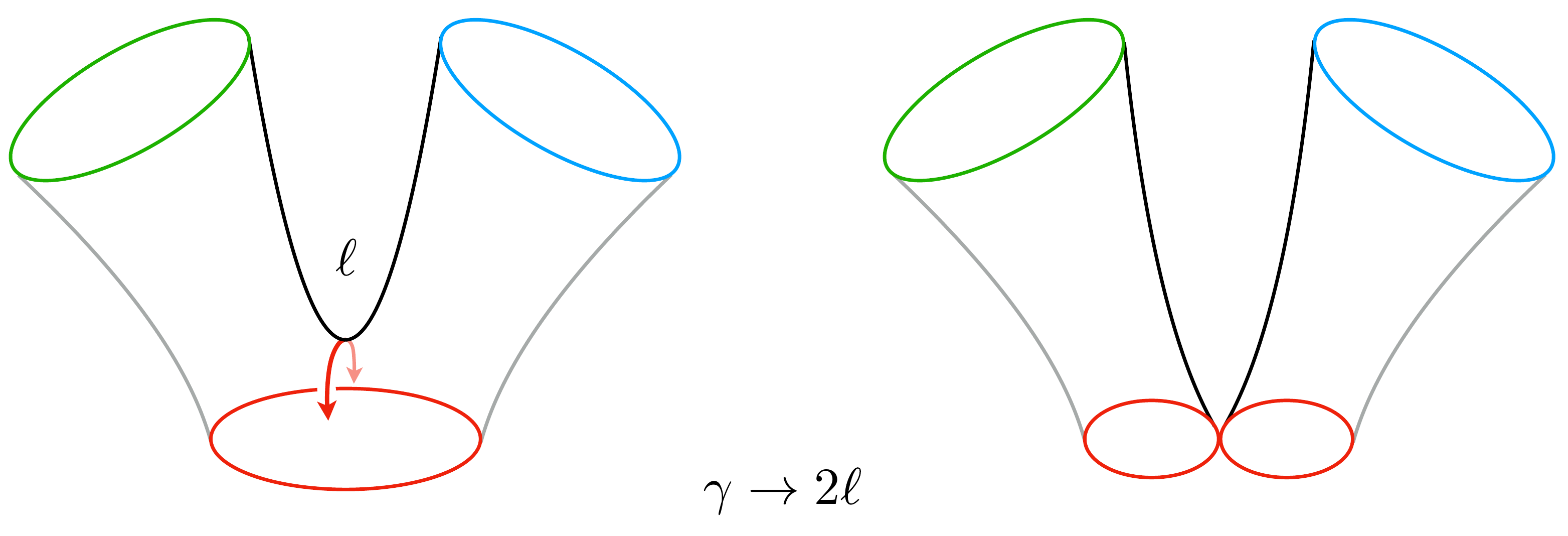}
\end{center}
\caption{Anomalous dimension $\gamma$ versus bridge length $\ell$. Increasing the energy $\gamma$ brings the worldsheet closer to the operator at the boundary by exciting energetic giant mirror magnons. The divergence occurs when the worldsheet reaches the boundary and crosses the double-trace process associated to the cut-opened bridge.}\label{Break} 
\end{figure}

The leading pole at $\beta = 0$ maps to a short-distance singularity on the worldsheet. As one can see from Eq.~(\ref{B1cut}) after performing the shift to the grading $\eta =  -1$, it signals the point where the one-magnon integral stops converging at $r \rightarrow \infty$. The phenomenon is thus driven on this side by giant mirror magnons carrying a very large energy. As said earlier, the specificity of these magnons is that they can reach the boundary of AdS. Heuristically, the giant magnons push the bridge towards the boundary where the worldsheet splits and the divergence occurs, see figure \ref{Break}. 

Formula (\ref{get}) also has zeros at specific positions. These ones relate to R-symmetry and supersymmetry. The former puts constraints on the $\su(4)$ weights $[q, p, q]$ of an operator falling in the OPE of two BPS operators, which must find room in the RHS of the tensor product, see, e.g, \cite{Eden:2001ec}
\beq\label{tensor}
[0, L_{1}, 0]\otimes [0, L_{2}, 0] = \bigoplus\limits_{n=0}^{L_{2}}\bigoplus\limits_{m=0}^{L_{2}-n} [m, L_{1}+L_{2}-2n-2m, m]\, ,
\eeq
where we used that $L_{1}\geqslant L_{2}$. Further constraints come from supersymmetry, when the operator belongs to a long multiplet, see~\cite{Eden:2001ec,Chicherin:2015edu,Dolan:2002zh,Dolan:2001tt,Dolan:2006ec}. In terms of the labels of the primary in the $\eta = -1$ grading, they read
\beq\label{n0}
n > 0\, , \qquad n+m < L_{2}\, ,
\eeq
and they shrink the triangular sum in (\ref{tensor}) on two sides. We can understand these inequalities as saying that not only should the $\eta = -1$ primary fit in the RHS of (\ref{tensor}) but also the superpartners listed in table \ref{primaries}. Operator (\ref{fields}) has length $L$ and scalar charge $Y$, i.e.~the $\su(4)$ labels $[Y, L-2Y, Y]$. Hence, $m = Y$ and $n = \ell_{B}$, and according to (\ref{n0}) we must have
\beq
\ell_{A}  -Y > 0\, , \qquad  \ell_{B} > 0\, ,
\eeq
where we used that $\ell_{A} + \ell_{B} = L_{2}$. Nicely, the denominator in (\ref{get}) kills the structure constant precisely when these conditions are not met.

We should stress that these arguments hold regardless of the strength of the coupling. Hence, poles and zeros should be observed at weak coupling as well. Yet, they are not as easy to see in this regime. For example, the poles and zeros in $\B$ are colliding when $\gamma \sim 0$ and, in particular, the leading zero at $\ell_{B}=0$ only becomes manifest after the neighbouring pole at $\ell_{B} = \tfrac{1}{2}\gamma$ has been handled properly. Disentangling this pair requires to analytically continue in $\ell_{B}$ and to resum terms in the weak coupling series which are enhanced when $\ell_{B}\sim \tfrac{1}{2}\gamma$. The analysis is done in Appendix \ref{weak} for completeness to leading order at weak coupling. In Appendix \ref{recurrence} we show that the pole persists at any $g$, in the simple case of scalar operators.

\subsection{Wrapping and string}

We will now discuss what happens for a classical length $L\sim g$ where one should in principle expect wrapping corrections to kick in. Recall first that for a classical length $\mathcal{L} = L/2g = O(1)$, the continuum of states breaks up into discrete energy levels. This follows from the quantization of the momenta,
\beq\label{free}
\sinh{\theta_{i}} = 2\pi n_{i}/\mathcal{L}\, ,
\eeq
with $n_{i} \in \mathbb{Z}_{\neq 0}$, and leads to the well-known BMN spectrum,
\beq\label{BMN}
\gamma = \sum_{i=1}^{M}(\sqrt{1+(\frac{2\pi n_i}{\mathcal{L}})^{2}}-1)\, .
\eeq
Besides this effect, which accounts for corrections that are power suppressed in $\mathcal{L}$, there are wrapping corrections. The latter come from virtual magnons going around the chain and are exponentially small in the case at hand \cite{Ambjorn:2005wa,Bajnok:2008bm}. A nice property of the BMN spectrum is that it is free from such corrections: Neither the energy formula (\ref{BMN}) nor the quantization conditions (\ref{free}) receive exponentially small additions. This is because the worldsheet theory is free at strong coupling. (There is not even a vacuum energy shift, because of supersymmetry.)

We will argue that something similar happens for the HHL structure constants of interest, meaning that one only needs to plug the quantized momenta in (\ref{get}) to keep track of the full dependence on $\mathcal{L}$, as long as the third bridge length $\ell_C  = L-\ell_{A}$ is much bigger than the dimension of the half-BPS operator $\mathcal{O}_{2}$.

To be more precise, let us write the structure constants in the form
\beq
C^{\circ\circ\bullet}/C^{\circ\circ\circ} = \mathcal{N}_{L}\mathcal{A}(\ell_{A})\mathcal{B}(\ell_{B}) \W_{\ell_{A}, \ell_{B}}(l_{C})\, ,
\eeq
where the first three factors are as before. The new factor $\mathcal{W}$ incorporates the wrapping corrections when $l_{C} = \ell_{C}/2g \sim \mathcal{L} \sim \mathcal{L}_{1}$ is held fixed. It drops out when $l_{C} \rightarrow \infty$,
\beq
\mathcal{W} \rightarrow 1\, ,
\eeq
since then nothing can pass through the bridge $C$. For $l_{C} = O(1)$, there are plenty of mirror magnons, with relativistic energies, going through, and one naively expects $\W$ to be a complicated function of all the quantum numbers in the problem, i.e., the splitting lengths $\ell_{A, B}$ and the Bethe roots. The claim is that $\mathcal{W}(l_{C}) = 1$ for any $\ell_{A, B} = O(1)$ and for any state with $\gamma = O(1)$.

As said earlier, we cannot rely entirely on the hexagon series for the wrapping corrections. To overcome this problem we will first extract information about them from the classical string analysis. This one requires that all lengths be classical. Hence, we shall take $l_{A, B, C} = \ell_{A, B, C}/2g = O(1)$ to begin with. Removing then the areas associated to the bridge $A$ and to the bridge $B$, we read from \cite{Kazama:2016cfl}
\beq\label{R}
\log{\W(l_{C}, l_{B}, l_{A})} = \mathsf{A}_{1}-\mathsf{A}_{2}+\mathsf{A}_{3}\, ,
\eeq
with 
\beq\label{Ai}
\mathsf{A}_{i} = \eta \int_{C^{+}} \frac{du(x)}{2\pi} \{2\textrm{Li}_{2} (e^{-l_{i}\epsilon(x)})-\textrm{Li}_{2} (q^{-\eta n_{i}}_{+}(x)e^{-l_{i}\epsilon(x)}) - \textrm{Li}_{2} (q^{-\eta n_{i}}_{-}(x)e^{-l_{i}\epsilon(x)})\}
\eeq
where $n_{i} = 1,2$ for $i$ odd, even and where $\eta = \pm 1$ refers to pure-$\Y$ and pure-$\D$ states. The formula only applies to the latter states in rank 1 sectors. The length $l_{i}$ takes three values,
\beq
l_{1} = l_{C}\, , \qquad l_{2} = l_{C} + l_{A} \, , \qquad l_{3} = l_{C}+l_{B}+l_{A} \, ,
\eeq
associated, respectively, to magnons on the bridge $C$, on the two bridges $C\cup A$, and on the three bridges $C\cup B \cup A$. The arguments of the dilogarithms are twisted using (\ref{qpm}) which are raised here to the power $\eta$ to account for the two families of states.

Next, to get an expression for energy $\gamma = O(1)$, we discretise the states and linearise the twists in (\ref{Ai}), as done previously for the $\B$ amplitude, see Section \ref{Sect3}. The areas in (\ref{R}) can then be replaced by coefficient $\d_{l}$, defined in Eq.~(\ref{logd}), and yield
\beq\label{Wlll}
\W(l_{C}, l_{B}, l_{A}) = \prod_{j=1}^{M} \frac{\d_{l_{C}+l_{A}}(\theta_{j})^2}{\d_{l_{C}}(\theta_{j})\d_{l_{C}+l_{B}+l_{A}}(\theta_{j})}\, .
\eeq
Note that the dependence on $\eta$ and thus on the flavours dropped out. We take it as evidence that the equation holds universally.

Similar (flavour-independent) combinations are found for the pp-wave Neumann coefficients. To be precise, given that the latter are associated to the near-collinear splitting of a string, one must take a length to be small classically. Taking e.g.~$l_{A}\rightarrow 0$ yields
\beq
 \B\,  \W = \prod_{j=1}^{M} \frac{\d_{l_{B}}(\theta_{j})\d_{l_{C}}(\theta_{j})}{\d_{l_{B}+l_{C}}(\theta_{j})}\, ,
\eeq
for the amputated structure constants,%
\footnote{According to \cite{Dobashi:2004nm,Dobashi:2004ka,Lee:2004cq} the pp-wave SFT vertex relates to amputated structure constants. We read it here as saying that one should remove the amplitude of the evanescent bridge.} which agrees with the length-dependent factor of the Neumann coefficients, see e.g.~\cite{Bajnok:2017mdf}, after replacing $b_{l}\rightarrow d_{l} = b_{l}|_{f\rightarrow 1}$. Note for the comparison that the bridge lengths can be identified in the SFT kinematics with the lengths of the operators.%
\footnote{E.g.~$l_{B} \sim \mathcal{L}_{2}, l_{C} \sim \mathcal{L}$ and $l_{B}+l_{C} \sim \mathcal{L}_{1}$ when $l_{A}\rightarrow 0$.}
One could proceed similarly for $l_{B} \rightarrow 0$ and find a similar agreement with the expressions in~\cite{Bajnok:2017mdf,Bajnok:2015hla}, removing this time the $\B$ amplitude.

Formula (\ref{Wlll}) shows that the wrapping factor is highly non-trivial when the three bridge lengths are comparable. However, it also predicts that there are no wrapping effects when one length is much larger than the others. We observe indeed that the $b$-coefficients in the numerator and in denominator cancel when $l_{A, B} \rightarrow 0$,
\beq\label{Wto1}
\lim\limits_{l_{A, B} \rightarrow 0}\W(l_{C}, l_{B}, l_{A}) \rightarrow 1\, , \qquad \forall l_{C}\, ,
\eeq
a feature which can be traced back to the fact that (\ref{R}) has no linear term around $q=1$ when $l_{1} = l_{2} = l_{3}$.

Taking the limit $l_{A,B}\rightarrow 0$ is not quite the same as setting $\ell_{A, B} = O(1)$. The mirror magnons leave the relativistic region and become giant when the lengths are finite. Hence, in order to complete the argument, we should allow for giant mirror magnons on the bridges $A$ and $B$ and study their interactions with the magnons sitting on the bridge $C$. The latter stick to the relativistic domain, since $\l_{C} \sim \mathcal{L}$ stays finite, meaning that there is a big energy gap between these magnons and those in $A$. For such disjoint kinematics, the hexagon formula should be free of wrapping divergences and we can turn to it to estimate the giant magnon effects.

The hexagon formula \cite{Basso:2015eqa} is predicting that the mirror magnons in $B$ decouple while those in $A$ and $C$ have mutual interactions controlled by
\beq
\prod_{u_i\in A, w_{j}\in C} \tilde{\Delta}_{a_{i}c_{j}}(u_{i}, w_{j})^{-1}\, ,
\eeq
with $\tilde{\Delta}$ as in (\ref{Delta}). However, these interactions go away, $\tilde{\Delta}\rightarrow 1$, in the case at hand, since the magnons in $C$ are relativistic. We conclude from it that the mutual interactions in $\W_{\ell_{A}, \ell_{B}}(l_{C})$ are localised on the relativistic modes. These modes are well described by the classical string analysis which predicts that they cancel out, see Eq.(\ref{Wto1}). In other words,
\beq
\W_{\ell_{A}, \ell_{B}}(l_{C}) = \lim\limits_{l_{A, B} \rightarrow 0}\W(l_{C}, l_{B}, l_{A}) \rightarrow 1\, ,
\eeq
which is the statement that the structure constants are free from wrappings for $l_{C} \simeq \mathcal{L} = O(1)$.

\section{Conclusion}\label{Concl}\label{Sect6}

In this paper we studied HHL structure constants at strong coupling using the hexagon framework. We found that the mirror sums describing these correlators simplify drastically and can be computed exactly for any bridge lengths. For finite bridges, we observed that the answer splits into a global factor, written in terms of Gamma functions, and an internal part, given by the product of the BMN energies. This factorisation mimics the separation between the two kinematical domains that contribute, namely the giant mirror magnons and the low-lying relativistic magnons.

We have seen that this factorisation is robust and applies to a large family of operators, containing derivatives and scalar fields. However, as broad as this family is, it does not include neutral pairs, like $\D\bar{\D}$ or $\Y\bar{\Y}$. It would be interesting to consider these pairs as they might enable a more precise comparison with the form factors found in the context of the SFT vertex, which are singlets under the $O(8)$ symmetry of the free worldsheet theory. (They might also be needed for applications to higher-point functions.) These pairs are associated to a deeper layer of auxiliary roots in the nested Bethe ansatz construction. They are described by more complicated transfer matrices and it is not obvious that the strategy followed in this paper will apply to them. Optimistically, the mirror sum analysis can be bypassed and the finite-bridge amplitudes be bootstrapped using analyticity, crossing, supersymmetry, etc.

It would also be nice to find the precise counterparts of all these amplitudes in the worldsheet theory. An operatorial definition might facilitate the calculation of more general HHL form factors and shed light on the absence of wrapping advocated in this paper. It may also help filling the gap with the near-flat space limit~\cite{Minahan:2012fh, Bargheer:2013faa, Minahan:2014usa}. One could perhaps reverse-engineer the formula obtained using integrability to reconstruct the worldsheet vertex.

Although we could not match our findings with direct worldsheet calculations, we observed that they have all the desired features to stand as a string correlators in $AdS_{5}\times S^{5}$. Notably, the singularities associated to the mixing with double-traces are packed inside Gamma functions, in agreement with general results from Witten diagrams.%
\footnote{Similar formulae were also found at weak coupling in the large spin limit \cite{Alday:2013cwa,Alday:2016mxe} using conformal bootstrap ideas. They hold when $\Gamma_{\textrm{cusp}} \sim 0$ for a flux-tube energy $\gamma \sim 2\Gamma_{\textrm{cusp}}\log{S} = O(1)$.}
The latter in fact predicts more Gamma functions than we have found here. The missing ones come with large arguments and drop out in the HHL kinematics. The semiclassical string formula is hinting at their ``reappearance" in the forms of ratios of $b$-coefficients, see Eq.~(\ref{Wlll}), which, as we have seen, are the classical counterparts of the Gamma functions. 

\begin{figure}
\begin{center}
\includegraphics[scale=0.35]{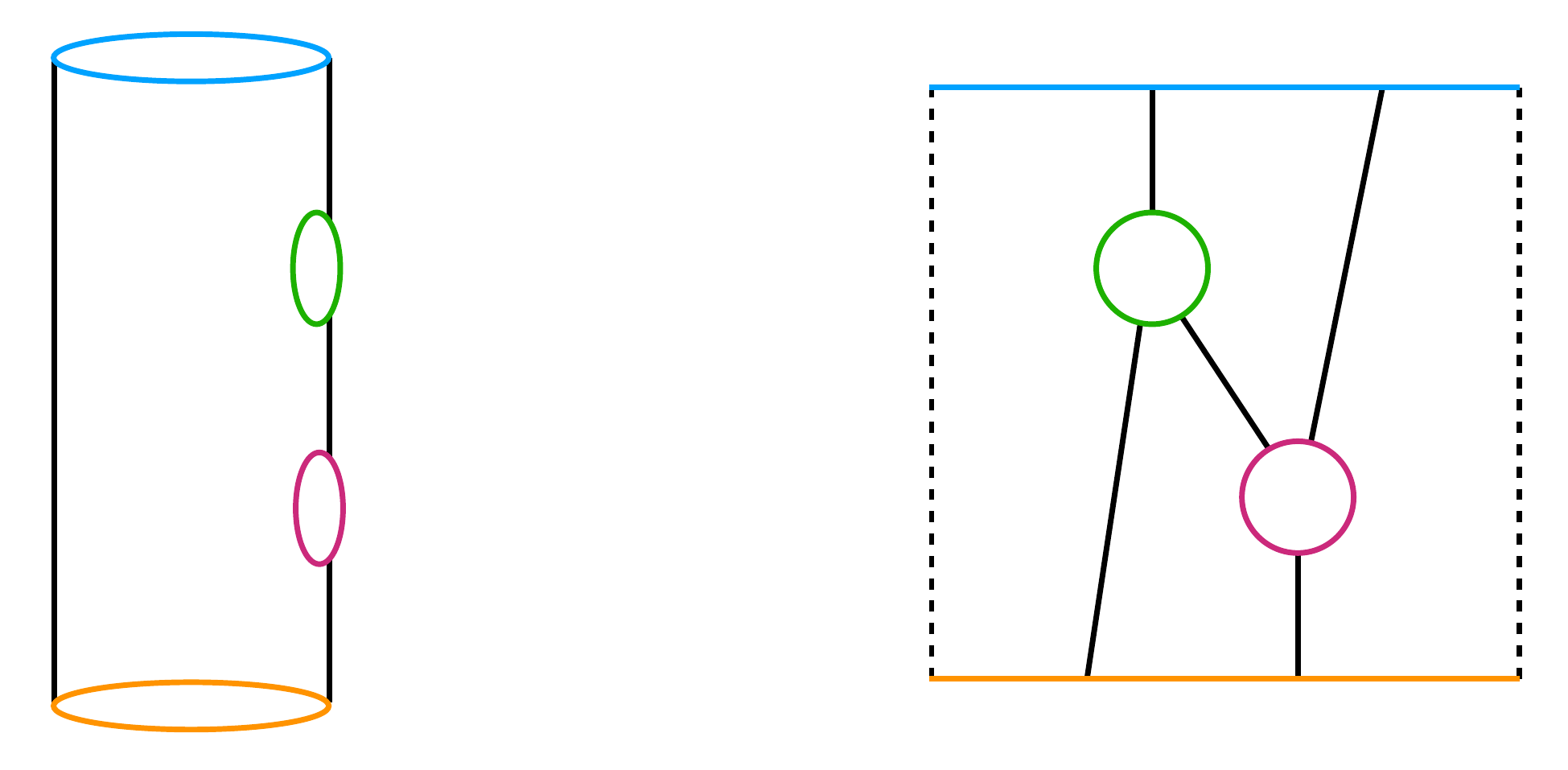}
\end{center}
\caption{Cartoon of a large-charge correlator with two light operators inserted on a large cylinder. Flattening the worldsheet and connecting the blobs with bridge lengths give the graph in the right panel, up to a flip. Up to double traces, the low-lying states flowing between the two insertions in the closed-string channel are BMN states like those studied in this paper.}\label{4pt} 
\end{figure}

Double-trace induced Gamma functions are omnipresent in holographic calculations and enter very naturally in the Mellin integrands of CFT/AdS correlators \cite{Mack:2009mi,Penedones:2010ue,Rastelli:2016nze}. It would be fascinating to establish a connection with the integrability formulae found here by considering large-charge correlators of the type shown in figure \ref{4pt}. These HHLL correlators can be obtained by sewing two pair-of-pants together using a complete sum of BMN operators, modulo double-traces. It is tempting to see in this inclusive sum the start of a Mellin integral. If so, the formulae obtained in this paper could help exploring the Mellin amplitudes in a more stringy regime.

It would also be interesting to compare our findings with the formulae obtained in \cite{Kostov:2019stn,Kostov:2019auq} for large-charge 4pt functions. Although both arise from hexagons the comparison is not immediate since they run with different transfer matrices, which is also why they describe different observables of the boundary theory. For the 4pt functions, the transfer matrix $\T$ is twisted such as to accommodate for the cross ratios \cite{Fleury:2016ykk}. This innocuous operation makes a difference for the strong coupling scaling, since then $\T$ is of order $O(1)$, as for semiclassical states, and the sum over bound states is regularised. It means that the saddle point will be trapped in the relativistic region, for generic values of the cross ratios. To escape from it and connect to our story, one might have to scale the cross ratios and work very close to the OPE limit / BPS point $\T = 0$. This is also suggested by the funny scaling $\sim g^{\frac{1}{2}\gamma}$ that we observed in this paper. It would be worth exploring this connection further in view of understanding how the 4pt function hexagon formula resums the tower of higher-rank structure constants, as well as to extract useful information about the latter if possible. It would also be interesting to  explore how the ``softening'' of the scaling with the coupling, which we observed here for the two-bridge amplitude, is realised in the hexagonalised 4pt functions. It is certainly hinting at the need to average over the various mirror channels. The latter operation proved to be important at weak coupling \cite{Fleury:2016ykk} to reproduce properties of the gauge-theory correlators. Similar magics might also be key at strong coupling to move away from the classical regime and match with the supergravity correlators.

\section*{Acknowledgments}

We thank Jo\~ao Caetano for collaboration at an early stage of this project. We also thank Romuald Janik and Miguel Paulos for discussions. B.B. acknowledges the Pauli Center for Theoretical Studies (ETH-Z\"urich) for the warm hospitality during the last stage of this project. This work was supported by the French National Agency for Research grant ANR-17-CE31-0001-02. The work of D.L.Z was supported in part by the European Research Council (Programme ERC-2012-AdG-320769 AdS-CFT-solvable).

\appendix

\section{Hexagon amplitude and transfer matrix}\label{App1}

The state-dependent factor in the hexagon integrals (\ref{Bn}) comprises an abelian piece and a matrix part. We can write them in more conventional terms using
\beq
\begin{aligned}
&\F_{a}(u) = \prod_{j=1}^{M}\frac{1/x^{[-a]}-x_j^{-}}{x^{[+a]}-x_{j}^{-}} h_{a1}(u^{3\gamma}, u_{j})\, , \\
& \T_{a}(u) = \prod_{j=1}^{M}\frac{x^{[+a]}-x_{j}^{-}}{1/x^{[-a]}-x_j^{-}} \times T_{a}(u^{3\gamma})\, ,
\end{aligned}
\eeq
with $h_{a1}(u, v)$ the dynamical part of the hexagon form factor between a bound state and a fundamental magnon \cite{Basso:2015zoa} and with $T_{a}(u)$ the forward $PSU(2|2)$ transfer matrix in the $a$-th antisymmetric representation \cite{Beisert:2006qh}. These two quantities are analytically continued to the mirror kinematics, by means of three mirror rotations $u\rightarrow u^{3\gamma}$, see e.g.~\cite{Basso:2015zoa} for the details of this transformation. We recall the general formulae for $\F_{a}$ and $\T_{a}$ below and relate them to the ones used in the bulk of the paper.

\subsection*{Hexagon amplitude}

Following \cite{Basso:2017muf} we can write the abelian part as
\beq\label{F-app}
\F_{a}(u) =  \prod_{i=1}^{M}F_{a1}(u, u_{i})\tilde{\sigma}_{a1}(u, u_{i})\, .
\eeq
Here, $\tilde{\sigma}_{a1}$ is closely related to the (fused) BES dressing phase; it admits a bi-linear expansion,
\beq\label{BESp}
\log{\tilde{\sigma}_{a1}}(u, v) = \sum_{n, m\geq 1} c_{2n, 2m+1}(\tilde{q}_{2n}q_{2m+1}-q_{2n}\tilde{q}_{2m+1})\, ,
\eeq
over a set of charges,
\beq
\tilde{q}_{r}(u) = \frac{1}{r-1}((x^{[+a]})^{1-r}+(x^{[-a]})^{1-r})\, , \qquad q_{r}(v) = \frac{i}{r-1}((y^{+})^{1-r}-(y^{-})^{1-r})\, ,
\eeq
with $y^{\pm} = x(v\pm i/2)$ and $x^{[\pm a]} = x(u\pm ia/2)$, and  with the BES coefficients \cite{Beisert:2006ez},
\beq\label{BESc}
c_{2n, 2m+1} = 2(-1)^{n+m} (2n-1)(2m)\int\limits_{0}^{\infty}\frac{dt}{t}\frac{J_{2n-1}(2gt)J_{2m}(2gt)}{e^{t}-1}\, ,
\eeq
with $J_{k}(z)$ the $k$-th Bessel function of the 1st kind. The other factor is given by
\beq\label{Fa1}
\begin{aligned}
\log{F_{a1}(u, v)} &= \sum_{m\geqslant 1}2(2m)(-1)^{m}\int\limits_{0}^{\infty}\frac{dt}{t} \frac{\sin{(ut)}e^{-at/2}J_{2m}(2gt)}{e^{t}-1} q_{2m+1}(v) \\
&+\sum_{m\geqslant 1}2(2m-1)(-1)^{m}\int\limits_{0}^{\infty}\frac{dt}{t}\frac{(\cos{(ut)}e^{-at/2}-J_{0}(2gt))J_{2m-1}(2gt)}{e^{t}-1}q_{2m}(v)\, .
\end{aligned}
\eeq

The large-momentum behaviour of mirror-magnon integrand is controlled by the asymptotic behaviour of the abelian factor, when
\beq
\sqrt{u^2+a^2/4} \rightarrow \infty\, ,
\eeq
with $u\sim a$. In this limit, the integrals in (\ref{Fa1}) receive dominant contributions from small $t$. All the integrals in the first line are power suppressed while the ones in the second line tend to constants, if not when $m = 1$. For $m = 1$ the integral scales logarithmically. Taking all of this into account, we find
\beq\label{large-r}
\log{\F_{a}(u)} = \gamma\log{\sqrt{u^2+a^2/4}} + \log{F(g)} + O(1/\sqrt{u^2+a^2/4})\, ,
\eeq
where we also used that the dressing phase (\ref{BESp}) is suppressed in this limit. Here, $F(g)$ is a state-dependent constant,
\beq\label{Fofg}
\log{F(g)} = \gamma\gamma_{E} + \sum_{m\geqslant 1}2(2m-1)(-1)^{m}\int\limits_{0}^{\infty}\frac{dt}{t}\frac{gt \delta_{m, 1}-J_{0}(2gt)J_{2m-1}(2gt)}{e^{t}-1}Q_{2m}\, ,
\eeq
and
\beq
Q_{r} = \sum_{j=1}^{M} q_{r}(u_{i})
\eeq
is the total charge of the Bethe state, with $\gamma = 2gQ_{2}$.

The strong coupling formula (\ref{shift}) is obtained by considering a mirror magnon with rapidities $x^{[\pm a]} = O(1)$ and a Bethe state with charges
\beq\label{charges}
Q_{r} \simeq \sum_{i = 1}^{M} \frac{x_i^{2-r}}{g(x_{i}^2-1)} = O(1/g)\, .
\eeq
The dressing phase (\ref{BESp}) reduces at strong coupling to the AFS phase \cite{Arutyunov:2004vx}
\beq
c_{2n, 2m+1} = g \delta_{n, m} - g\delta_{n-1, m}\, .
\eeq
This follows immediately from (\ref{BESc}) after rescaling the integration variable $t\rightarrow t/2g$, expanding at large $g$ and using known integrals for products of two Bessel functions.
Plugging these coefficients into (\ref{BESp}) and using (\ref{charges}) one obtains
\beq\label{dress-str}
\prod_{i=1}^{M}\tilde{\sigma}_{a1}(u, u_{i}) = \prod_{i=1}^{M}(1-\frac{1}{x^{[+a]}x_{i}})(1-\frac{1}{x^{[-a]}x_{i}})\,  \exp{\{gQ_{1}\frac{x^{[+a]}+x^{[-a]}}{x^{[+a]}x^{[-a]}}\}}\, .
\eeq
We proceed similarly for (\ref{Fa1}), taking the strong coupling limit, with $u^{[\pm a]}/2g = (u\pm ia/2)/2g$ fixed, and applying
\beq\label{fBessel}
\int\limits_{0}^{\infty}\frac{dt}{t^2} e^{\pm iu^{[\pm a]}t/2g} J_{k} (t) = \frac{1}{2k}\{\frac{1}{k+1}(\pm i/x^{[\pm a]})^{k+1}+\frac{1}{k-1} (\pm i/x^{[\pm a]})^{k-1}\}\, .
\eeq
The $u$ independent term in the second line of Eq.~(\ref{Fa1}) can be dropped for $k\neq 1$, since $\int_{0}^{\infty} dt J_{0}(t)J_{k}(t)/t^2 = 0$, for $k$ odd. One must be more careful for $k=1$, as in this case the $u$ independent term is needed for removing the small $t$ divergence in the LHS of (\ref{fBessel}). This subtraction happens to be equivalent to expanding the RHS around $k=1$ and discarding the polar part $\sim 1/(k-1)$. Straightforward algebra gives then
\beq
\begin{aligned}
\prod_{j=1}^{M}F_{a1}(u, u_{j}) = (x^{[+a]}x^{[-a]})^{gQ_{2}}\prod_{i=1}^{M}\tilde{\sigma}_{a1}(u, u_{i})^{-1}\, ,
\end{aligned}
\eeq
or equivalently $\mathcal{F}_{a}(u) = (x^{[+a]}x^{[-a]})^{\frac{1}{2}\gamma}$. This result is consistent with the asymptotic formula (\ref{large-r}), given that $F(g) \rightarrow g^{-\gamma}$ at strong coupling. 

\subsection*{Transfer matrix}

The general formula for the eigenvalue of the transfer matrix can be found in \cite{Beisert:2006qh}. Performing three mirror rotations and fixing the normalisation appropriately, we find
\beq
\begin{aligned}\label{T-app}
\T_{a}(u) = &\prod_{i=1}^{N}\frac{1}{(1/x^{[-a]}-y_{j})(1-1/x^{[+a]}y_{j})}\\
&\,\,\,\times \big[ t_{a+1}(u) - t_{a}(u^{+}) \f^{-}_{a}(u) - t_{a}(u^{-})\f^{+}_{a}(u) + t_{a-1}(u) \f^{+}_{a}(u)\f^{-}_{a}(u)\big]\, ,
\end{aligned}
\eeq
with $u^{\pm} = u\pm i/2$. Here, the $y$'s are fermionic roots at the first nested level in the $\eta = +1$ grading. The $f$'s are functions of the main roots only,
\beq \label{eqn-fpm}
f^{\pm}_{a}(u) = \prod_{j=1}^{M} \xi_{j}^{\mp 1} \frac{1-1/x^{[\pm a]}x_{j}^{\mp}}{1-1/x^{[\pm a]}x_{j}^{\pm}}\, ,
\eeq
with $\xi_{j} = (x^{+}_{j}/x^{-}_{j})^{1/2}$ and with $\prod_{j=1}^{M}\xi_{j} = 1$ for a cyclic state. $t_{a}(u)$ is the eigenvalue of a $\sl(2)$ XXX transfer matrix with auxiliary spin $\tfrac{1}{2}(a-1)$. For a state in the $\psu(1,1|2)$ sector, we get to consider its vacuum eigenvalue with the $y$'s acting as inhomogeneities. It reads
\beq\label{ta}
t_{a}(u) = \sum_{k= 0}^{a-1} \Qy(u^{[a-1-2k]})\, ,
\eeq
with $\Qy$ the Baxter polynomial for the fermionic roots, which we recall here for convenience,
\beq\label{P}
\Qy(u) = g^{-N}\prod_{i=1}^{N} (u-v_{i}) = \prod_{i=1}^{N}(x-y_{i})(1-1/xy_{i})\, ,
\eeq
with $u = g(x+1/x)$ and $v_{i} = g(y_{i}+1/y_{i})$. These formulae are written assuming that the roots $\{y_{i}\}$ are non vanishing. The case $y=0$ is obtained as a limit and leads to the relation (\ref{y0}).

We can group the four terms of the transfer matrix differently, as done in Eq.~(\ref{Ta1}), by introducing
\beq\label{Tpm}
\T^{\pm} =\Q^{\pm} \,  (1 - f^{\pm}_{a}(u))\, ,
\eeq
with $\Q^{\pm}$ as in Eq.~(\ref{calQpm}), and
\beq\label{T0}
\T^{\, 0} = \prod_{j=1}^{N}\frac{1}{(1/x^{[-a]}-y_{j})(1-1/x^{[+a]}y_{j})}t_{a-1}(u) \left(1-f^{+}_{a}(u)\right) \left(1-f^{-}_{a}(u) \right)\, .
\eeq
This recasting relies on
\beq
t_{a+1}(u) - t_{a}(u^{+}) - t_{a}(u^{-}) + t_{a-1}(u) = 0\, ,
\eeq
and on other simple recurrence relations, which all follow from the definition (\ref{ta}).

At strong coupling, when the roots $\{x_{j}\}$ are of order $O(1)$, the terms in brackets in (\ref{T0}) and (\ref{Tpm}) are of order $O(1/g)$. Plugging (\ref{eqn-xpmLargeG}) into (\ref{eqn-fpm}) and expanding in $1/g$, we get
\beq
1 - f^{\pm}_{a}(u) \simeq \pm \frac{i}{g} \sum_{j=1}^{M}\frac{x_{j}}{(x_{j}^2-1)(x^{[\pm a]}x_{j}-1)} \, ,
\eeq
leading to the expressions (\ref{calTpm}) and (\ref{calT0}).

Yet another regime where $f\sim 1$ is weak coupling. Then $x^{[\pm a]} \sim u^{[\pm a]}/g$ and
\beq
f^{\pm}_{a}(u) = 1\mp \frac{i\gamma}{2 u^{[\pm a]}} + O(g^{4})\, .
\eeq
We have similarly that $y_{i}\sim v_{i}/g$ and therefore,
\beq\label{wcT}
\T^{\pm} \simeq \pm \frac{i\gamma P(u^{[\mp a]})}{2 u^{[\pm a]} P(0)}\, , \qquad \T^{\, 0} = O(g^4)\, ,
\eeq
after using (\ref{calQpm}) and~(\ref{P}).

\section{Analytic regularisation}\label{reg-app}

In this appendix we evaluate the one-particle integral at strong coupling using a different regularisation. Namely, instead of introducing a hard cut-off for the IR and UV regions, we modify the behaviour of the measure close to $r^2 = x^{[+a]}x^{[-a]} \sim 1$, using
\beq
\tilde{\mu}_{a} (u) \rightarrow \tilde{\mu}_{a} (u) (x^{[+a]} x^{[-a]}-1)^{\alpha}\, ,
\eeq
where $\alpha \sim 0$, and compute the integrals in the two regions (over a complete domain in each case).

In the IR region we keep $a = O(1)$ and expand the integrand for $g\rightarrow \infty$. The additional $\alpha$-dependent factor yields
\beq
(x^{[+a]} x^{[-a]}-1)^{\alpha} = (e^{\E_{a}}-1)^{\alpha} \sim \left(\frac{a\varepsilon(s)}{2g}\right)^{\alpha}\, , 
\eeq
using $\E_{a} \sim a \varepsilon(s)/2g$, with $s$ along the unit circle. The sum over $a$ gives Riemann zeta function evaluated at $1-\alpha$,
\beq
\sum\limits_{a=1}^{\infty} a^{-1+\alpha} = \zeta(1-\alpha) \simeq -\frac{1}{\alpha} + \gamma_{E} \, ,
\eeq
and thus
\beq
\B_{1}^{IR} = \int \frac{du(s)}{2\pi}  \T(s) \zeta(1-\alpha)\left(\frac{\varepsilon(s)}{2g}\right)^{\alpha} \simeq \int\frac{du(s)}{2\pi}\T(s)  (-\frac{1}{\alpha} + \gamma_{E} + \log{(\varepsilon (s)/2g)})\, ,
\eeq
with domain of integration $\{|s| = 1, \Im m (s) >0\}$. The integral is easily taken and yields
\beq\label{BI}
\B_{1}^{IR} = -\frac{\gamma}{2\alpha}+\frac{\gamma\gamma_{E}}{2} + \log{C}\, ,
\eeq
with $C$ as in (\ref{Cst}).

For the UV region, we take $a\sim g$ and replace the sum $\sum_{a}$ by an integral. Performing the average over the angular variable, we obtain (\ref{B1cut}) integrated from $1$ to $\infty$ with the deformed weight,
\beq\label{BII}
\B_{1}^{UV} = \int\limits_{1}^{\infty} \frac{\gamma dr}{r^{2\b+1}(r^2-1)^{1-\alpha}} = \frac{\gamma \Gamma(1+\b-\alpha)\Gamma(\alpha)}{2\Gamma(1+\b)} = \frac{\gamma}{2\alpha} -\frac{\gamma}{2}(\gamma_{E}+\psi(1+\b)) + O(\alpha)\, .
\eeq
(Note that $\B^{UV}$ is defined for $\alpha>0$ and then continued, while $\B^{IR}$ was defined for $\alpha <0$ and then continued.)

Now, both $\B^{IR}$ and $\B^{UV}$ are singular when $\alpha \rightarrow 0$, but their poles readily cancel in the sum. We get
\beq
\B_{1} = \B_{1}^{IR} + \B_{1}^{UV} = \log{C} -\frac{\gamma}{2}\psi(1+\b)\, ,
\eeq
in agreement with the result obtained using a hard cut-off.

\section{Near-extremal correlators at weak coupling}\label{weak}

In this appendix we analyse the octagon amplitude $\B(\ell)$ at weak coupling. First, recall the weak coupling scaling of the $m$-particle integral $\B_{m}$ at weak coupling \cite{Basso:2015eqa,Eden:2018vug}
\beq
g^{2m(m+\ell)} \times \T^{m}
\eeq
where $\T$ is for the eigenvalue of the transfer matrix. In the following we shall study what happens when we analytically continue the integrals close to $\ell = -1$ for a primary in the $\eta = +1$ grading. This point is the turning point for the exponent of the one-particle integral. The higher-$m$ integrals are still parametrically smaller and can be discarded in a first approximation. Hence, in the following we shall restrict our discussion to the vacuum and one-particle integral
\beq\label{B01}
\B(\ell) \simeq 1+\B_{1}(\ell) \, .
\eeq

\begin{figure}
\begin{center}
\includegraphics[scale=0.3]{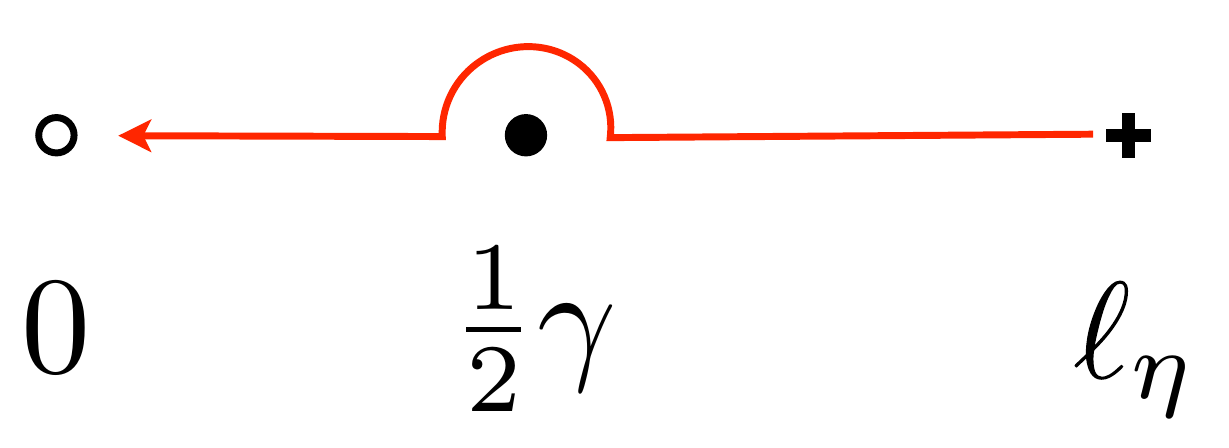}
\end{center}
\caption{The one-particle integral is well defined for any $\ell_{\eta} > 0$ to any order at weak coupling, with $\ell_{\eta} = \ell+\tfrac{1}{2}(1+\eta)$. However, the weak coupling series only converges for $\ell_{\eta} > \tfrac{1}{2}\gamma$. To see what happens for $\ell_{\eta} = 0$ one must analytically continue around the pole at $\ell_{\eta} = \tfrac{1}{2}\gamma$. To leading order at weak coupling, the residue at the pole determines the value of the integral at $\ell_{\eta}=0$.}
\end{figure}

To begin with, let us consider a scalar state and expand all the ingredients at weak coupling. We get
\beq
\T_{a}(u) \simeq \frac{a\gamma}{2(u^2+a^2/4)}\, , \qquad \F_{a}(u) \simeq 1 \,, \qquad \tilde{\mu}_{a}(u) \simeq \frac{a g^2}{(u^2+a^2/4)^2}\, ,
\eeq
and, after combining all factors together,
\beq
\B_{1}(\ell) = \frac{\gamma}{2} \sum_{a\geqslant 1} \int \frac{du}{2\pi}\frac{a^2 g^{2+2\ell}}{(u^2+a^2/4)^{\b+3}}\, ,
\eeq
where here $\beta = \ell$. The integral over $u$ is easily taken and the sum is expressed in terms of the Riemann zeta function,
\beq\label{B1ell}
\B_{1}(\ell) = \frac{\gamma 2^{3+2\b}g^{2+2\ell}\Gamma(\frac{5}{2}+\b) \zeta(3+2\b)}{\Gamma(\tfrac{1}{2})\Gamma(3+\b)}\, .
\eeq
It is smooth for $\ell > -1$ and, despite the many factors, analytically continued to a function with a single pole at $\beta = -1$, coming from the Riemann zeta function,
\beq\label{B1pole}
\B_{1}(\ell) \sim  \frac{\tfrac{1}{2}\gamma g^{2+2\ell}}{\beta+1}\, .
\eeq
The singularity comes from the large $r = \sqrt{u^2+a^2/4}$ behaviour of the integrand. We can verify it by converting the sum over $a$ into an integral, as done earlier at strong coupling,
\beq
\sum_{a} \int \frac{du}{2\pi} \rightarrow  \frac{1}{2}\int\limits_{1}^{\infty}dr\, \oint \frac{ds}{2\pi i s}\, ,
\eeq
with the polar coordinates $u^{[\pm a]} = r s^{\pm 1}$ and with a cut off set at $r=1$ for convenience. (Since we are only interested in the residue at the leading pole, only the large $r$ behaviour matters and the precise value of the lower bound is irrelevant. We also symmetrised the domain of integration in $a$, using the parity symmetry of the integrand.) Straightforward algebra gives
\beq
\B_{1}(\ell) \sim -\tfrac{1}{2} \gamma g^{2(1+\ell)}\int\limits_{1}^{\infty} \frac{dr}{r^{3+2\b}} \oint \frac{ds}{2\pi is} (s-1/s)^2 = \frac{\gamma g^{2(1+\ell)}}{2(1+\b)}\, ,
\eeq
in agreement with (\ref{B1pole}).

If we were to expand further the integrand at weak coupling we would find higher poles at $\beta = -1$. They originate from the logarithms in the $\F$-factor and can be resummed using the asymptotic behaviour (\ref{large-r}), that is
\beq
\F_{a}(u) \sim F(g) (u^2+\frac{a^2}{4})^{\tfrac{1}{2}\gamma}\, ,
\eeq
with $F(g) = 1+O(g^2)$. The rest is power suppressed at large $r$ or just modifies the overall factor by subleading corrections in $g^2$. Hence, resumming the leading singularities is done by using $\beta = \ell-\tfrac{1}{2}\gamma$ in the formulae above.

In particular, we find that the singular part of the loop corrections reads
\beq
\B_{1}(\ell) = \frac{\tfrac{1}{2}\gamma g^{2(1+\ell)}(1+O(g^2))}{1+\ell-\tfrac{1}{2}\gamma} \, .
\eeq
This behaviour determines the result at $\ell = -1$ to leading order at weak coupling. It gives $\B_{1}(\ell = -1) = -1$ which cancels the tree-level part in (\ref{B01}). Hence, as expected, the amplitude vanishes at $\ell = -1$. (Checking the zero at higher loops would require to take into account multi-particle corrections at some point, as well as to expand further the one-particle integral around $\beta = -1$.)

This analysis generalises to a generic primary in the $\psu(1,1|2)$ sector with $\Qy(0)\neq 0$, after replacing the transfer matrix by its general expression, see Eq.~(\ref{wcT}),
\beq
\T_{a}(u) \simeq \frac{i\gamma}{2} \frac{u^{[-a]}\Qy(u^{[-a]})-u^{[+a]}\Qy(u^{[+a]})}{u^{[+a]}u^{[-a]}\Qy(0)}\, .
\eeq
It is saying that adding $y$'s reduces a priori the degree of convergency of the integral, since the numerator is then of a higher degree in both $u$ and $a$. Nonetheless, we can still evaluate the integral for large enough $\beta$. For a generic $P$, we use
\beq
\sum\limits_{a=1}^{\infty} a\int \frac{du}{2\pi} \frac{(u\pm ia/2)^{2k+1}}{(u^2+a^2/4)^{3+\b}} = \pm i\frac{4^{2+\b-k}\Gamma(2k-\b-1)\Gamma(\tfrac{5}{2}+\b-k)}{2\Gamma(\tfrac{1}{2})\Gamma(k-\b-1)\Gamma(3+\b)} \zeta(3+2\b-2k)\, ,
\eeq
together with the fact that only odd powers survive. The large $a$ behaviour (i.e., power counting) of the sum in the LHS directly maps to the pole of the $\zeta$-function in the RHS. We note however that this pole at $\b = k-1$ is absent from the full result when $k\neq 0$, since it multiplies a zero of the polynomial $\Gamma(2k-\b-1)/\Gamma(k-\b-1)$. The sole exception is $k=0$, which brings us back to (\ref{B1ell}). Therefore, we find here again a single, simple pole at $\beta = -1$ with the same residue as before. It predicts a zero at $\ell = -1$. We can verify all of that using the angular average, which returns
\beq
\oint  \frac{ds}{2\pi i s} a\T_{a}(u) =\frac{\gamma}{2r}\oint \frac{ds}{2\pi is} (s-1/s) \frac{(r/s)\Qy(r/s)-(rs)\Qy(rs)}{\Qy(0)} = \gamma\, .
\eeq
for any polynomial $\Qy$ such that $P(0)\neq 0$.

The bottom line is that the position of the pole and its residue are independent of the spin $N$, despite the fact that the integral is seemingly less convergent at higher $N$. It agrees with the analysis in \cite{Basso:2015zoa} which shows that the convergency of the integral improves and becomes spin independent if one first takes the sum over $a$. We see the same phenomenon here as coming from the angular integration.

Note finally that the state independence of the pole is unlikely to hold beyond the $\psu(1,1|2)$ sector. The next layers of magnons will add conjugate derivatives $\bar{\D}^n$, or equivalently boxes, which increase the twist of the operator, $t = L+\gamma +2n$. One thus expects the leading pole to show up at $\ell = -1+ n+\tfrac{1}{2}\gamma$, with a small residue $\sim g^{2(1+\ell)} \sim g^{2n}$.

\section{Leading pole at finite coupling}\label{recurrence}

In this appendix we study the leading pole of the amplitude $\B(\ell)$ at finite coupling and its relation to the one-particle integral $\B_{1}(\ell)$. We consider a scalar state for simplicity. The asymptotic behaviour of the one-magnon integrand for large $u$ and large $a$ is then simply given by
\beq
e^{-\ell \tilde{E}_{a}(u)}\tilde{\mu}_{a}(u) \F_{a}(u) \T_{a}(u) \sim  \frac{a^2g^{2(1+\ell)} F(g) \gamma (1+\tfrac{1}{2}\gamma)}{2(u^{2}+a^2/4)^{1+\b}}\, ,
\eeq
with $\b = 1+\ell - \tfrac{1}{2}\gamma$. We used here the asymptotic behaviour~(\ref{large-r}) for $\F_{a}(u)$ and the fact that $\T_{a}(u) \sim a\gamma (1+\tfrac{1}{2}\gamma)/(2u^{[+a]}u^{[-a]})$ which holds for scalar states.
It implies, after integrating over $a$ and $r$, that the leading pole in $\B_{1}$ is given by
\beq\label{B1sing}
\B^{sing}_{1} = \frac{\tfrac{1}{2}\gamma (1+\tfrac{1}{2}\gamma) g^{\gamma}F(g)}{1+\ell-\tfrac{1}{2}\gamma}\, ,
\eeq
with $F(g)$ the state-dependent constant (\ref{Fofg}). In principle, this behaviour could be shifted by the higher contributions in $\B(\ell)$. This is not the case because of the interaction (\ref{Delta}). It is such that sending a mirror magnon to infinity increases the bridge length for its companions, since
\beq
\lim\limits_{x^{[\pm a]}\rightarrow \infty} \tilde{\Delta}_{ab}(u, v) = (y^{[+b]}y^{[-b]})^{-2}\, .
\eeq
In other words, the pole is triggered by a single mirror magnon, which decouples from the rest up to a length shift. Namely,
\beq
\B (\ell) \sim \B^{sing}_{1}  \times \B(\ell+2)\, ,
\eeq
close to the pole, or equivalently
\beq
\mathtt{Res}_{\ell \, = \, -1 +\frac{1}{2}\gamma}\,  \B(\ell) = (1+\ell)(2+\ell) g^{\gamma}F(g) \B(\ell+2)\, ,
\eeq
using (\ref{B1sing}). This equation should hold at any coupling $g$. It is verified at weak coupling, see Appendix \ref{weak}, and at strong coupling, using (\ref{oct1}) together with $F(g) \rightarrow g^{-\gamma} $. What is more, in the latter case, a recurrence relation holds away from the pole,
\beq
\B(\ell+2)/\B(\ell)|_{g\rightarrow \infty} = \frac{(1+\ell - \tfrac{1}{2}\gamma)(2+\ell - \tfrac{1}{2}\gamma)}{(1+\ell)(2+\ell)}\, ,
\eeq
according to Eq.~(\ref{oct1}).

\bibliography{biblio}
\bibliographystyle{JHEP} 

\end{document}